# Minute-long quantum coherence enabled by electrical depletion of magnetic noise

Cyrus Zeledon[1], Benjamin Pingault[2,3,1], Jonathan C. Marcks[2,3,1], Mykyta Onizhuk[1], Yeghishe Tsaturyan[1], Yu-Xin Wang[4], Benjamin S. Soloway[1], Hiroshi Abe[5], Misagh Ghezellou[6], Jawad Ul-Hassan[6], Takeshi Ohshima[5,7], Nguyen T. Son[6], F. Joseph Heremans[2,3,1], Giulia Galli[1], Christopher P. Anderson[8], and David D. Awschalom[1,2,3,9,*]

[1]Pritzker School of Molecular Engineering, University of Chicago, Chicago, Illinois 60637, USA
[2]Materials Science Division, Argonne National Laboratory, Lemont, IL 60439, USA
[3]Q-NEXT, Argonne National Laboratory, Lemont, IL 60439, USA
[4]Joint Center for Quantum Information and Computer Science, University of Maryland, College Park, MD 20742, USA
[5]National Institutes for Quantum and Radiological Science and Technology, 1233 Watanuki, Takasaki, Gunma 370-1292, Japan
[6]Department of Physics, Chemistry and Biology, Linköping University, SE-581 83 Linköping, Sweden
[7]Department of Materials Science, Tohoku University, Aoba, Sendai, Miyagi 980-8579, Japan
[8]Department of Materials Science and Engineering, University of Illinois Urbana-Champaign, Urbana, IL 61801, USA
[9]Department of Physics, University of Chicago, Chicago, Illinois 60637, USA

*Corresponding author. Email: awsch@uchicago.edu

**ABSTRACT.** Integrating solid-state spin defects into classical electronic devices can enable new opportunities for quantum information processing that benefit from existing semiconductor technology. Here, we investigate the impact of bias control of an isotopically purified silicon carbide (SiC) p-i-n diode on the coherence of embedded spins. We show that the diode allows for the depletion of not only the electrical, but also the magnetic noise sources. This results in extended relaxation and coherence times of individual electronic and nuclear spins, with Hahn echo times exceeding values reported for single spins in any platform (> 100 seconds). These results demonstrate the importance of materials control and electronic device integration to create highly coherent solid-state quantum technology.

## I. INTRODUCTION

Optically-active electron spins in the solid state are promising platforms for quantum communication [1] and quantum sensing [2–5] because of their spin-photon interface [6, 7], long coherence times [8–10], high-fidelity single-shot readout [8, 11], and access to nuclear spins as long-lived quantum memories [12, 13]. These systems have enabled sensing at the nanoscale [14, 15] and formed the backbone of nascent quantum networks [16, 17]. However, the uncontrolled electric and magnetic noise from defects and impurities in the solid-state host limits the spin coherence of these systems [18, 19], hindering many of these applications. Specifically, magnetic noise can arise from paramagnetic impurities and naturally occurring nuclear spin isotopes, and electric noise from impurity atoms and lattice vacancies acting as charge traps. To remedy this situation, several approaches have been explored to improve the host matrix. These include isotopic purification [12, 20, 21] and growth with low precursor and chamber contamination to reduce paramagnetic impurities [19, 22, 23] to limit magnetic noise. They also include semiconductor device engineering [22, 24, 25], optimized growth conditions [22, 23, 25], and annealing out of host crystal vacancies [21, 23, 25, 26] to address electric noise. Unfortunately, these approaches do not fully eliminate all known noise sources and thus generally need to be complemented by Hamiltonian engineering on the spin qubit to decouple from the remaining noise [6, 27–29]. Such decoupling sequences can, however, be complex to implement and interfere with the realization of quantum protocols. Here, we demonstrate how a diode structure can in-situ deplete both electric and magnetic noise sources in silicon carbide (SiC), beyond the limit of materials growth optimization, and achieve record-long nuclear spin Hahn echo coherence times, reflecting the noise reduction in the engineered device.

SiC is a semiconductor widely used in the high-power electronics industry with wafer-scale production and excellent doping capabilities. Its large bandgap and naturally low nuclear spin content also make it an excellent host for optically active spin qubits for quantum technologies [30]. Among these spin qubits, the neutral divacancy ($VV^0$) [31], consisting of neighboring silicon and carbon vacancies, has attracted attention due to its optically interfaced electron spin (S = 1), making it a promising building block for quantum networks [12, 22, 32, 33] and quantum sensing [34, 35]. However, despite mature high purity SiC production capabilities, including isotopic purification [12, 36], impurities and defects still limit the optical and spin properties of SiC spin qubits [12, 24]. While qubit optical linewidths can be reduced and emission wavelengths stabilized through embedding into a diode structure [24], further improvements of spin properties have remained elusive.

Our approach leverages high quality materials combined with a diode structure aimed at further reducing the impact of impurities and defects on the spin properties of $VV^0$. Having minimized the nuclear spin bath through isotopic engineering, we show that operating the diode in the depletion regime not only improves optical properties (an electric dipole moment) but also extends the coherence of the $VV^0$ electron spin (a magnetic dipole moment), evidence that the diode can reduce electric noise as well as magnetic noise. This is confirmed by double-quantum Ramsey [37–41] measurements and by the extension of the population decay time ($T_1$) of a neighboring nuclear spin. Through coherent control of this nuclear spin, we show that a combination of diode-induced magnetic noise depletion and the frozen core effect from the $VV^0$ allows the Hahn echo coherence time of the nuclear spin to exceed a minute, to date the longest reported value across any qubit platform, including solid-state, atoms, and ions [42–50]. This not only furthers our understanding of the impact of noise sources on solid-state qubits but also demonstrates how simultaneous active electric and magnetic engineering in a device can significantly improve solid-state qubit performance and functionality.

## II. RESULTS

We first identify a single axial $VV^0$ (PL2) [31] located in the intrinsic region of a p-i-n diode in custom-grown isotopically purified 4H-SiC at 5 K (Fig. 1(a)). We characterize the electrical behavior of the diode and measure the corresponding Stark shift of the optical emission of the $VV^0$ under the effect of the local electric field in the diode (Fig. 1(b)). In forward bias, $VV^0$ experiences a Stark shift of several GHz from 0 to 120 V, but the rapid increase in current through the diode makes this regime unsuitable for stable operation. In reverse bias, we observe a minimal optical line shift between 0 V and -40 V, corresponding to the formation of a depletion region across the diode until this depletion region reaches the location of the defect [24]. Beyond this voltage, $VV^0$ emission is Stark-tuned over several GHz without significant leakage current through the diode (tens of nA), allowing for long term stability of the emission; see Appendix B. As previously reported [24], the diode bias voltage also impacts the optical linewidth (Fig. 1(c)). While the ~80 MHz linewidth measured at zero bias remains approximately constant in forward bias, it rapidly decreases down to 18(1) MHz in the depletion regime for reverse bias voltages below -50 V. This value is close to the expected natural linewidth of 11 MHz, corresponding to the measured optical lifetime of 15(1) ns; see Appendix B. This is explained by a reduction in electrical noise near the defect from fluctuating charge traps [24]. As the ground and excited states have a very similar g-factor [24], the optical linewidth of the $VV^0$ center does not depend on magnetic noise.

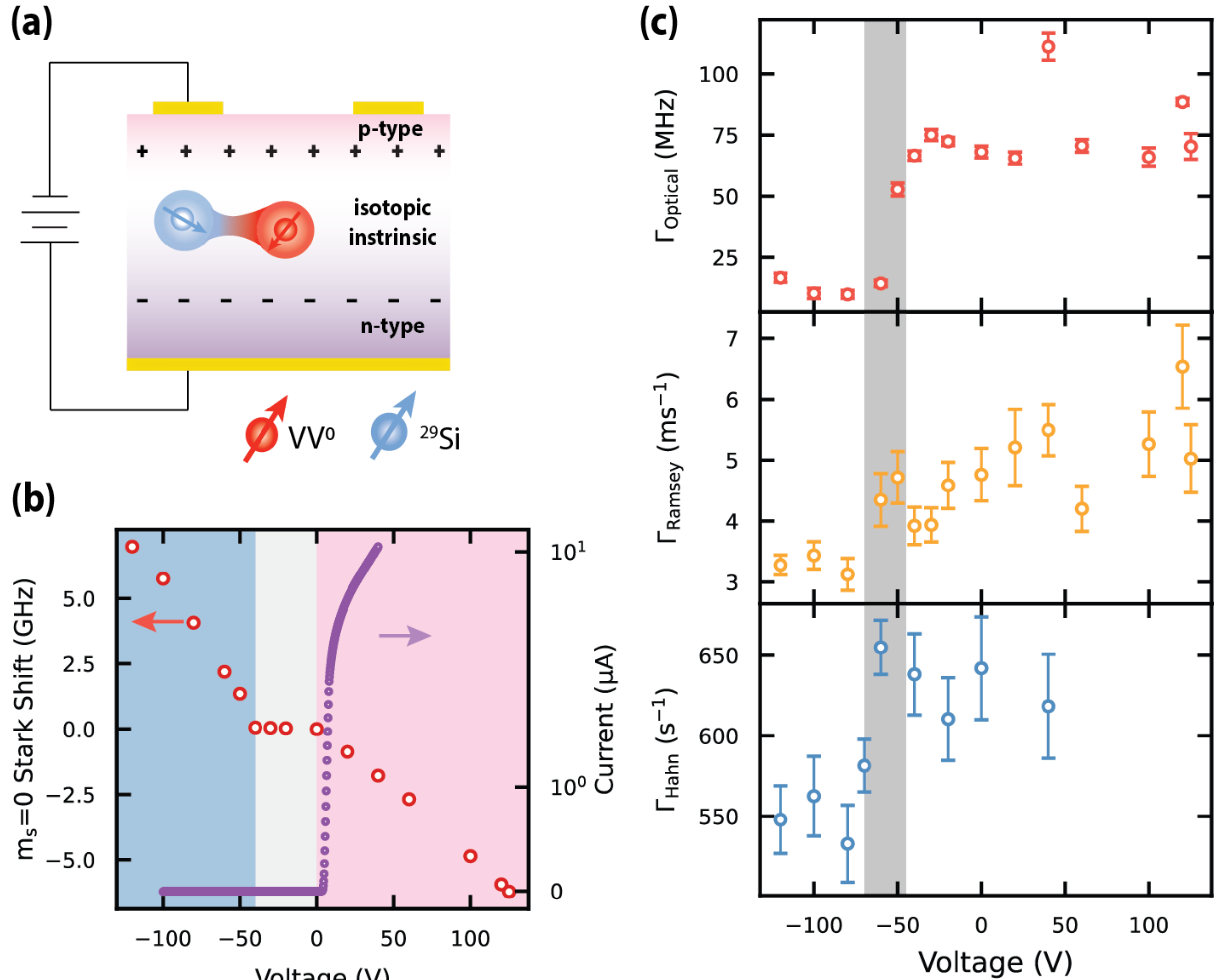

**FIG 1.** p-i-n diode and tuning the embedded $VV^0$ environment. (a) Schematic of diode structure showing the spins studied. (b) Current-voltage (I-V) curve (purple) of the diode and Stark Shift (red) of the $VV^0$ optical emission measured through resonant photoluminescence excitation. The three different colored panels show depletion, and forward (reverse) bias Stark shifts with minimal optical shifts (left to right). (c) Optical linewidth of the divacancy's $E_x$ transition (top), spin Ramsey dephasing rate (middle), and Hahn echo dephasing rate (bottom) as a function of applied bias. The shaded region is a guide for the eye to show the onset of depletion at the $VV^0$, and the vertical dashed line indicates the bias voltage at which subsequent measurements are performed. All data are taken at a magnetic field of 232 Gauss aligned with the $VV^0$ axis and at T = 5 K.

### A. Diode-dependent behavior of an isolated single divacancy spin

We then explore the impact of the diode bias voltage on the $VV^0$ electron spin coherence by measuring the Ramsey and Hahn echo dephasing rates between the $m_s$ = 0 and $m_s$ = -1 states at different applied voltages (Fig. 1(c)). Similar to the behavior observed for the optical linewidth, a forward bias leaves the dephasing rates relatively unchanged. However, when entering the depletion regime in reverse bias, below approximately -50 V, both dephasing rates decrease. The Ramsey coherence time reaches 320(30) μs — an improvement by approximately 50% compared to the unbiased value of 210(20) μs — which is the longest reported for electronic spins in SiC without using a clock transition [6, 12, 36]. The Hahn echo coherence time similarly improves by about 25%, reaching 1.9(1) ms. This highlights the diode's ability to modify the underlying dominant noise sources.

To determine the physical origin of these improvements, we perform double-quantum (DQ) Ramsey measurements, corresponding to a Ramsey sequence between the $m_s$ = -1 and $m_s$ = +1 states of $VV^0$. This technique removes common mode electric field noise but doubles the sensitivity to magnetic noise [37–41]. By performing these measurements at 0 V and -80 V (see Fig. 2(a)), we obtain DQ $T_2^*$ of 115(4) μs and 149(7) μs respectively, corresponding to approximately half the values measured with the single-quantum Ramsey sequence at the same bias voltages; see Appendix H. This indicates that the dephasing is dominated by magnetic fluctuations, and as a consequence, that the coherence improvement observed in the depletion regime of the diode can be ascribed to a reduction in magnetic noise [37–41].

Previous studies in isotopically engineered 4H-SiC material demonstrated Hahn echo coherence times of 2.3 ms for $VV^0$ and 0.8 ms for $V_{Si}$, which are lower than nuclear spin-limited theoretical values (~tens of ms for both defects [12, 36]) given the level of isotopic purification. Previous works attributed this limitation to cracks in the material from processing [36] and unaccounted for paramagnetic spins that decohere the controlled electron spins in the material faster than expected [12]. While it has been shown that depletion improves uncontrolled fluctuating electric fields [24], we here show that it can also deplete unwanted paramagnetic noise sources. Possible such sources include spin-containing charge traps, such as silicon vacancies ($V_{Si}$), nitrogen impurities, and carbon vacancies ($V_C$), which can be converted to a spin-less charge state [51], thereby improving the coherence of the $VV^0$ electronic spin; see Appendix H. The comparable improvements in the Hahn echo and Ramsey values suggest that we have reduced sources of both slow (~Hz) and fast (~kHz) magnetic noise; see Appendix.

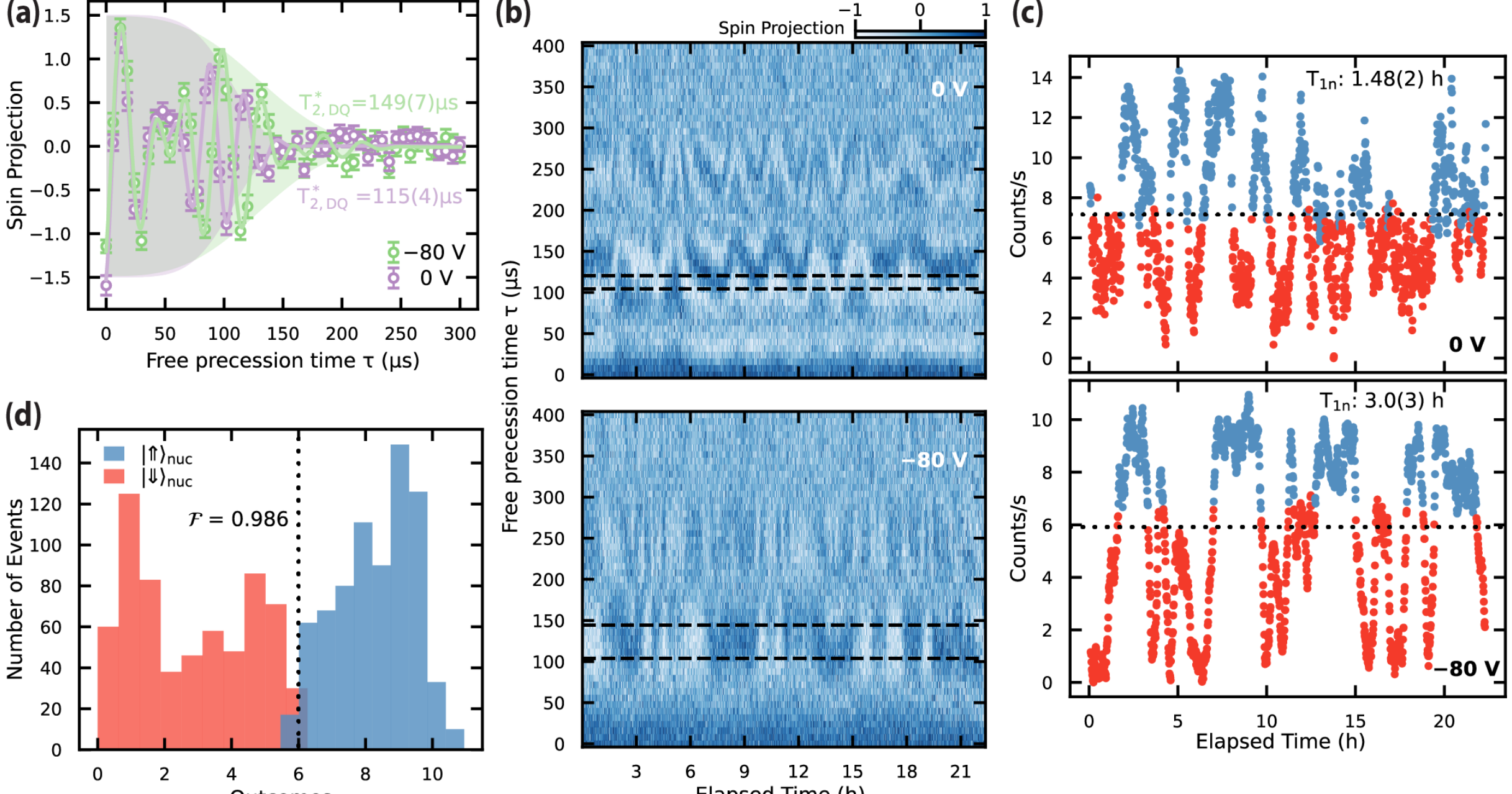

**FIG 2.** Understanding the noise and spin-related effects. (a) Double quantum Ramsey interference of a single $VV^0$ at 0 V and -80 V in reverse bias. (b) (top) A 22.5 hour Ramsey measurement scan of the $m_s = 0$ to $m_s = -1$ transition of the $VV^0$ while under 0 V of bias showing slow variations of the $VV^0$ Ramsey frequency throughout the scan. (bottom) A 22.5 hour Ramsey measurement scan of the $m_s = 0$ to $m_s = -1$ transition of the $VV^0$ while under -80 V of reverse bias showing stable $VV^0$ spin frequency. A detuning of 10 kHz is used for both scans. Integration time per point is 1 s and each Ramsey scan takes 57.5 s on average. The dashed lines in both plots bound the regions for the rolling average line cut in (c). (c) Shifted differential of the 0 and -1 fluorescence time trace showing single-shot readout of the nuclear spin and corresponding quantum jumps. The red is the nuclear down state, and the blue is the nuclear up state. The dashed lines are the thresholds that maximize readout fidelity (see the Appendix F and Ref. [42]). (d) Histogram of the counts from the repetitive readout in (c) demonstrating the ability to distinguish between the two nuclear spin states.

### B. Understanding slow noise

We then perform a Ramsey measurement on the $VV^0$ for 22.5 hours to understand the dynamics of the slow noise for both 0 and -80 V in reverse bias (Fig. 2(b)). In both experiments, we observe characteristic shifts in the Ramsey frequency that can be ascribed to flips of a single neighboring nuclear spin, as described in Ref. [42]. This is a result of the specific isotopic engineering of our sample, with 99.85% $^{28}$Si and 99.98% $^{12}$C, tuned to obtain approximately one nuclear spin with ~10 kHz coupling per $VV^0$, on average [12]. This design aims to balance the reduction in nuclear spin bath magnetic noise to maximize spin coherence, while maintaining (nominally one) accessible nuclear spin memory. We also observe that at 0 V there are continuous variations in the Ramsey oscillation frequency, indicative of slow shifts in local magnetic field noise. On the other hand, in depletion (-80 V), the frequency is much more stable (see Appendix G), and the shifts in Ramsey frequency are more abrupt, indicating that coupling to the single nuclear spin is the dominant effect. By measuring the Ramsey signal at a delay corresponding to the hyperfine coupling (see values below) [42] and monitoring it over the course of the 22.5 hours (Fig. 2(c)), we measure the nuclear spin state in a quantum non-demolition (QND) manner. This allows us to measure direct flips of the single nuclear spin and extract its relaxation time $T_1$. We observe that the diode impacts this relaxation time, increasing it by a factor of 2 in the depletion regime. This further confirms the reduction in magnetic noise evidenced in the electronic double-quantum Ramsey measurements (Fig. 2(c)). A histogram of QND measurements of the nuclear spin in the depletion regime is shown in Fig. 2(d), demonstrating single-shot detection of the nuclear spin state with a fidelity of 98.6%.

### C. Coherent control and coherence of a single $^{29}$Si nuclear spin

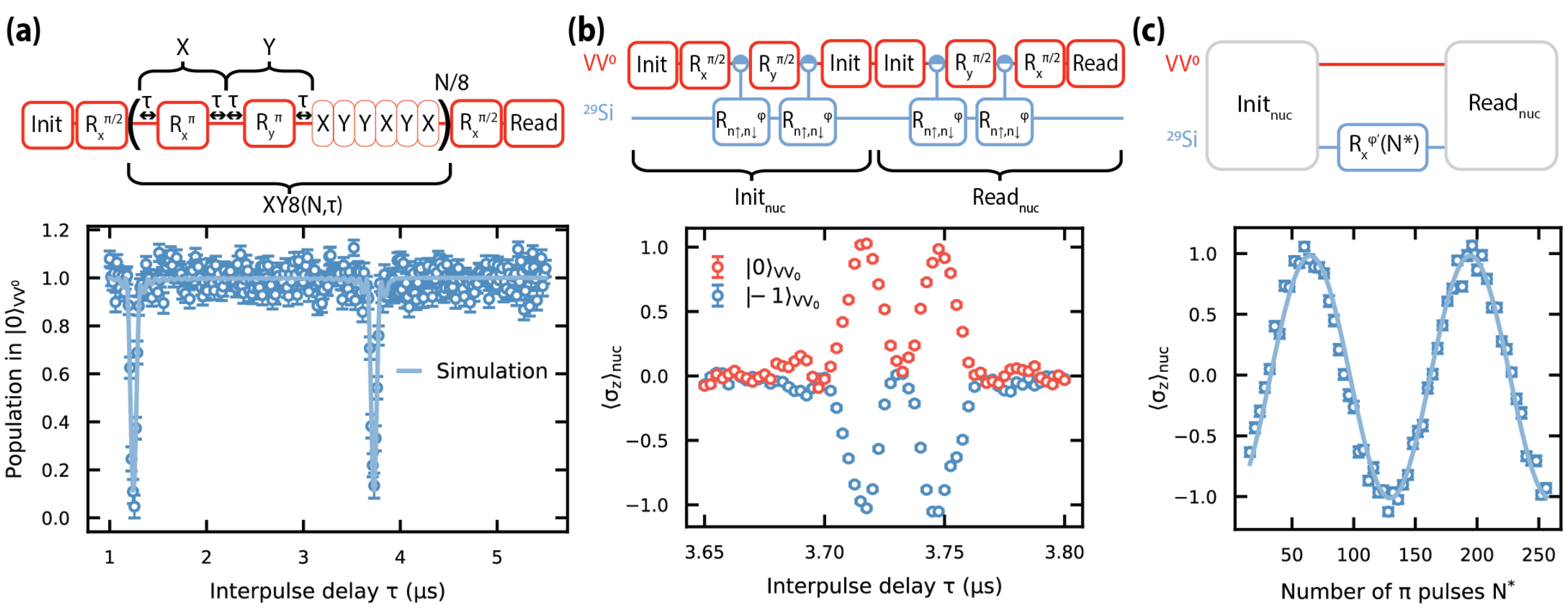

**FIG 3.** Coherent control of a single $^{29}$Si nuclear spin. (a) Pulse sequence for nuclear spin spectroscopy using XY8-N dynamical decoupling sequence on $VV^0$ (top), with N the total number of pulses. The rotation ($R_{x/y}{}^{\theta}$) gates represent rotation of the electron spin by an angle θ around the given axis of the Bloch sphere, and "Init" and "Read" refer to the optical initialization and readout of the electron spin state. The measurement is of the final $VV^0$ spin population in $|0\rangle$ using an XY8-32 sequence as a function of the interpulse time τ (bottom). The blue circles represent the experimental data, and the light blue line represents a simulation allowing the extraction of the hyperfine coupling parameters $\{A_{\parallel} = -8.1$ kHz and $A_{\perp} = 9.4$ kHz$\}$. (b) Pulse sequence for the nuclear spin initialization and readout gates (top). The conditional rotations are executed with XY8 sequences on the $VV^0$ spin. The measured values of the $VV^0$ spin projection after the nuclear initialization and readout gates are presented as a function of interpulse delay τ for different initial spin states of $VV^0$. The optimal delay time for highest initialization and readout fidelity is τ = 3.715 μs. (c) Pulse sequence (top) and measurement (bottom) for coherent control of the single nuclear spin by performing dynamical decoupling on the $VV^0$ spin with N* pulses, and the measured Rabi oscillations of the nuclear spin. Light blue line represents a sine fit over the data. $\langle\sigma_Z\rangle$ is the nuclear spin projection measured with the SCC readout on the $VV^0$ spin normalized to the maximum SCC contrast.

We then move on to controlling this nuclear spin. To do so, we first perform dynamical decoupling spectroscopy. By applying an XY8 pulse sequence on the $VV^0$ spin and varying the interpulse delay, we identify periodic resonances in the coherence function when the interpulse time is set to $\tau_k \approx (2k+1)\pi / (2\omega_l+A_{\parallel})$, where k is the integer order of the resonance dip, $\omega_l$ is the Larmor frequency of the target nuclear spin, and $A_{\parallel}$ is the longitudinal hyperfine coupling between the two spins (Fig. 3(a)). The signal can be fitted by simulating the spin dynamics, allowing the extraction of the hyperfine coupling parameters — $A_{\parallel}$ = -8.1(1) kHz and $A_{\perp}$ = 9.4(1) kHz, later confirmed through Ramsey measurements on the nuclear spin and in line with expectations from the isotopic engineering (see Appendix A) — and of the nuclear Larmor frequency, which is consistent with a single $^{29}$Si nuclear spin; see Appendix C. Previous results demonstrated nanoscale NMR and oscillations of weakly coupled nuclei in isotopically purified SiC [12], but there are no measurements to date of the initialization, control, and coherence properties of these potential quantum registers. Having identified the target nuclear spin, we proceed to initialize it and read it out with the pulse sequence presented in Fig. 3(b), which consists of a series of dynamical decoupling pulses on the electron spin. Based on the state of the $VV^0$ spin at the beginning of the sequence, the nuclear spin can be initialized in either the up or down state, for an appropriate interpulse delay, as shown in Fig. 3(b). By adding an extra dynamical decoupling sequence between the nuclear initialization and readout and varying the number of pulses on the $VV^0$ spin, we coherently drive Rabi oscillations of the nuclear spin (Fig. 3(c)) [52, 53].

Utilizing this control over the nuclear spin, we perform a Ramsey sequence on the $^{29}$Si nuclear spin to measure its coherence properties in the diode. Spin-to-charge conversion (SCC) readout of the $VV^0$ spin is implemented for these long nuclear spin coherence measurements to enhance the signal-to-noise ratio [8]. First, we perform Ramsey measurements with the $VV^0$ spin in the $m_s$ = 0 state during the free precession and with the diode unbiased, yielding a $T_2^*$ coherence time of 380(30) ms (Fig. 4(a)). Biasing the diode into the depletion regime at -80 V extends the $T_2^*$ coherence time to 520(10) ms. We therefore find that the charge depletion induced by the diode affects not only the electron spin but the nuclear spin as well, with a ~40% improvement in dephasing time, similar to the improvement seen in the electron spin $T_2^*$ of ~50%. It is indeed expected that the same paramagnetic noise that limits the $VV^0$ would also affect the nuclear spins in the material. In addition, we find that the coherence times are affected by the frozen core effect when the electron spin state is prepared in $m_s$ = -1. This occurs because the $VV^0$ electron spin induces a hyperfine field with a strong gradient on nearby spins that suppresses the transfer of polarization between them, benefiting the coherence time of the target spin [54]. In this situation, we measure a nuclear spin $T_2^*$ of 690(40) ms while in the depletion regime (Fig. 4(a)). It is interesting to note that the observed enhancement factor in the nuclear spin Ramsey time versus its $T_1$ relaxation time follows a simple scaling relation of $T_2^*(-80V)/T_2^*(0V) \simeq \sqrt{T_1(-80V)/T_1(0V)}$. Such a scaling law agrees with the microscopic spin-bath model where both nuclear Ramsey dephasing and relaxation are due to the same ensemble of weakly interacting magnetic impurities [13, 55, 76]; see Appendix I.

Finally, we perform a Hahn echo sequence on the single $^{29}$Si spin. We measure a coherence time of 10.2(6) s when the electron spin is prepared in the $m_s$ = 0 state (Fig. 4(b)). When the electron is instead prepared in $m_s$ = -1, the frozen core effect dramatically increases this coherence time. Despite only observing a small Hahn echo decay in Fig. 4(b), we extract the nuclear Hahn echo time $T_2$ with a $\chi^2$ goodness of fit test and find a lower bound of 105 s, assuming a stretch exponent of n = 2 typically measured in comparable quasi-static environments in diamond [43, 50], with 99% confidence (discussion in the Appendix D). We confirm these observations using second-order CCE simulations that predict that, due to the frozen core, Hahn echo decay times can be up to minutes long in our magnetic and charge noise-depleted environment; see Appendix I. Our measurements therefore demonstrate the importance of using electric fields to enhance both optical and spin properties of solid-state qubits, enabling minute-long coherence times for a single nuclear spin with a simple Hahn echo sequence. This is the longest reported value to date across all platforms, including those with a clock transition [42–49], and is on par with more complex systems with a decoherence protected subspace [50].

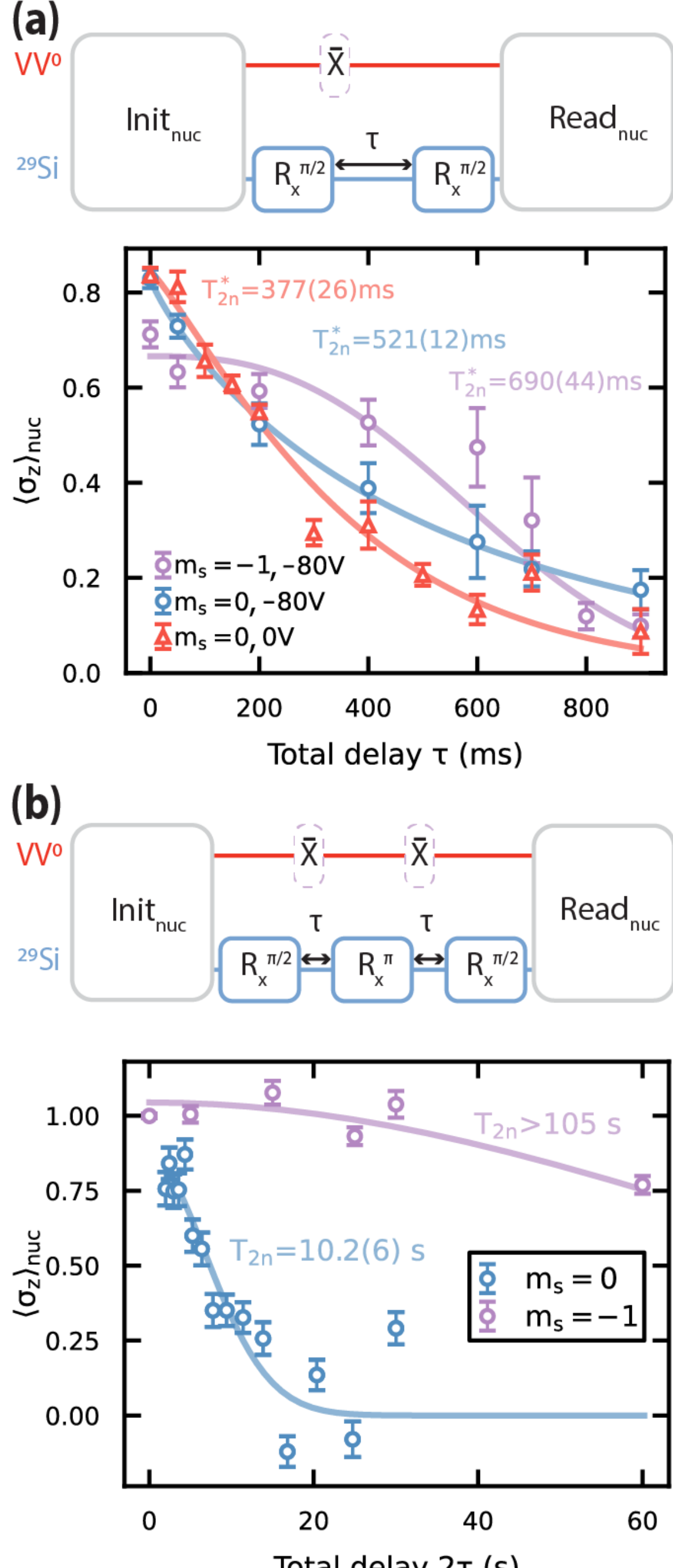

**FIG 4.** Coherence of a single $^{29}$Si nuclear spin. (a) Pulse sequence and bias voltage-dependent measurement of Ramsey interferometry on the nuclear spin when the electron is prepared in $m_s = 0$ (light red for 0 V bias and light blue for -80 V) and $m_s = -1$ (light purple), where pulses are the same as in Fig. 3(c). The Ramsey interference data in a narrow time window at each time point are fitted, and the amplitudes and associated errors then plotted. (b) Hahn echo pulse sequence and measurement of the nuclear spin under -80 V bias when the electron is prepared in $m_s = 0$ (light blue) and $m_s = -1$ (light purple). In the pulse sequence, the dashed purple boxes denote an optical initialization followed by a π pulse to set the VV$^0$ spin to the $m_s = -1$ state. $\langle \sigma_Z \rangle$ is the nuclear spin projection measured with the SCC readout on the VV$^0$ spin normalized to the maximum SCC contrast.

## III. DISCUSSION

We attribute these improvements across optical and spin measurements in SiC to the simultaneous reduction in electric and magnetic noise. While the reduction of electric noise is expected, the reduction of magnetic field noise upon reverse bias of the diode is confirmed by both the increased DQ Ramsey $T_2^*$ on the VV$^0$ and improvements on the lifetime and coherence of the $^{29}$Si nuclear spin ($S = 1/2$), which are only sensitive to magnetic noise. This reduction is hypothesized to arise from the depletion of paramagnetic charge traps near the defect and other affected spin species. Materials grown via chemical vapor deposition generally have defect densities at or below $10^{14}$ cm$^{-3}$ [12, 24, 36] causing a fundamental limit in the coherences of spin qubits. Of the likely paramagnetic defects, we expect that the amount of $V_{Si}$ from irradiation and growth will be negligible after annealing steps [56], leaving both nitrogen and $V_C$ as potential sources of magnetic noise. They have an electronic spin $S = 1/2$ in the -1 and 0 charge states respectively (which occur with a mid-gap Fermi level) [57–60]. Therefore, we hypothesize that the reduction in magnetic noise is due to nitrogen impurities and $V_C$ being converted into spinless positive charge states by a combination of electrical bias and photoexcitation [57–60, 76]. These results demonstrate that electric and magnetic noise in solid-state matrices can be mitigated in-situ with a simple semiconductor diode to improve the optical and spin properties of embedded qubits beyond materials optimization.

## IV. CONCLUSIONS AND OUTLOOK

The electrical depletion demonstrated here reduces both electric and magnetic noise from the host matrix. It thus improves optical linewidths, extends the nuclear spin lifetime, and extends coherence of both electron and nuclear spins to record values [42-50, 61]. We find an improvement to the electron spin coherence by more than 50% of non-depleted values, and minute-long nuclear spin Hahn echo coherence, exceeding values reported in other spin and atomic/ionic qubit platforms. These extended coherence times are ideal for quantum information protocols that require long-lived spins that can serve as memory qubits, while reducing the need for dynamical decoupling sequences or specific clock operation points. Fundamentally, this reflects our ability to engineer environmental noise in situ for quantum states to maintain their coherence, even in imperfect solid-state materials. The presented depletion technique can be readily applied to other qubits in semiconductors that are limited by electric field fluctuations [21, 62, 63] or paramagnetic sources [41, 62, 63]. Such a device could further be integrated with photonic structures, including cavities and waveguides [32, 64], where it could help preserve optical and spin properties that generally degrade in nanostructures due to surface charges. This would form the basis of a quantum node to develop a SiC-based quantum network, leveraging the industrial maturity of this platform. Other avenues that can be explored include memory-enhanced sensing where the register can increase the sensitivity and performance of the sensor [65, 66]. Overall, the diode-embedded spin platform presented here demonstrates the promise of integrating classical and quantum optoelectronics into highly coherent wafer-scale quantum technology.

## ACKNOWLEDGMENTS

We thank Dr. Leah Weiss, Grant T. Smith, Marquis M. McMillan, Dr. Joseph P. Blanton, Swathi Chandrika, Sanskriti Chitransh, Dr. Jacob S. Feder, Prof. Peter Maurer, and Dr. Conor Bradley for helpful experimental suggestions and discussions, Jered Feldman, William Hyland, and Julia Krueger for helpful considerations for the device processing.

This work was supported by AFOSR FA9550-23-1-0330 (C.Z., D.D.A.); AFOSR FA9550-22-1-0370 (C.Z., D.D.A., G.G.); Boeing through the Chicago Quantum Exchange (C.Z.); NSF ECCS-2442352 (C.P.A.); the U.S. Department of Energy, Office of Science, Basic Energy Sciences, Materials Sciences and Engineering Division through Argonne National Laboratory under Contract No. DE-AC02-06CH11357 (J.C.M. and F.J.H.); the U.S. Department of Energy Office of Science National Quantum Information Science Research Centers as part of the Q-NEXT center (B.P.); Laboratory Directed Research and Development (LDRD) funding from Argonne National Laboratory, provided by the Director, Office of Science, of the U.S. Department of Energy under Contract No. DE-AC02-06CH11357 (B.P.); NSF QuBBE QLCI (NSF OMA-2121044) (B.S.S.); Internationalisation Fellowships (No. CF20-0475) from the Carlsberg Foundation (Y.T.); the QuICS Hartree Postdoctoral Fellowship (Y.-X.W.); KAKENHI (20H00355) (T.O.); the Swedish Research Council (VR 2020-05444) (J.U.H.); the European Union's HORIZON project SPINUS (101135699) (J.U.H.); Vinnova (2024-00461) (N.T.S.); Horizon Europe projects QRC-4-ESP (101129663) and QUEST (101156088) (N.T.S.); the Knut and Alice Wallenberg Foundation (KAW 2018.0071) (J.U.H., N.T.S.); and Horizon Europe project QuSPARC (101186889) (J.U.H., N.T.S.). This work made use of the UChicago MRSEC (NSF DMR-2011854) and the Pritzker Nanofabrication Facility, part of the Pritzker School of Molecular Engineering at the University of Chicago, which receives support from the Soft and Hybrid Nanotechnology Experimental (SHyNE) Resource (NSF ECCS-2025633), a node of the National Science Foundation's National Nanotechnology Coordinated Infrastructure [RRID: SCR_022955].

## AUTHOR CONTRIBUTIONS

Conceptualization: C.Z., D.D.A. Experimental measurement: C.Z., B.P. Device fabrication: C.Z., Y.T. Data analysis: C.Z., B.P., J.C.M., C.P.A. Experimental setup development: C.Z., C.P.A., B.P., J.C.M., B.S.S. Theoretical understanding: Y.-X.W., M.O. Theoretical calculations: M.O. Electron irradiation: H.A., T.O. SiC p-i-n design and epitaxial growth: J.U.-H., M.G. Theoretical supervision: G.G. Supervision: D.D.A. Writing (review and editing): all authors contributed to manuscript preparation.

## COMPETING INTERESTS

The authors declare no competing interests.

## DATA AVAILABILITY

The data underlying this study will be available at Zenodo (TBD).

# APPENDIX

## APPENDIX A: Materials and Methods

### *1. Experimental Details*

All measurements are performed in a home-built confocal microscope using a closed-cycle Montana Cryostation s100 cryostat with a 0.85 numerical aperture 100x near-infrared objective at T = 5 K. All measurements are performed at a magnetic field of around 232 G (permanent Nd disk magnet) on a single PL2 $VV^0$ in 4H-SiC. We use a Thorlabs single-stage TEC resistive heater wrapped around the magnet with a LakeShore model 332 cryogenic temperature controller for a built-in PID to stabilize the temperature within less than a percent at 297 K. The magnet is boxed within the optical enclosure with foam for extra temperature stability. For narrow-line laser control and photoluminescence excitation (PLE) scans of our defect, we use two Toptica DLC PRO lasers. The collection of $VV^0$ emission is done through a 1060XP fiber and detected with a Quantum Opus Superconducting Nanowire Single Photon Detector (SNSPD). For charge reset/initialization of $VV^0$, we use a 705-nm laser for 30 ms at a laser power of ~200 nW at the sample. Spin initialization and readout are performed from each Toptica laser by pumping on the $A_1$ transition and the $E_x$ transition, respectively. For sequences involving coherences of the nuclear spin, we implement the spin-to-charge conversion readout as in Ref. [8]. A 1151-nm ionization laser is used from QPhotonics QFLD-1160-300S. Manipulation of the spin sublevels $m_s = 0$ to $m_s = -1$ ($m_s = 0$ to $m_s = +1$) of the electron is performed using 0.697-GHz (1.947-GHz) microwaves from an SRS SG396 applied through a draped, insulated Au wire above the sample surface. Microwave filtering and extinction are implemented to minimize pulse errors on the electron spin. We use a 185-MHz high pass filter directly after the signal generator and MW amplification and ZASWA-2-50DR+ microwave pulse switches (one before and one after the amplifier) for extinction and noise improvement. For high frequency measurements involving the electron spin, we use an Amplifier Research 25GS1G4A amplifier; for low frequency measurements for the electron and nuclear measurements, we use an Amplifier Research 30W1000B. We use IQ modulation of the source to control the phase of the microwaves to perform XY8 pulse sequences for the weakly coupled nuclear spin measurements. For all MW pulse sequences, we use Hermitian-shaped pulses [43] with π pulses of approximate length 1-2 μs in case of electron coupling to strongly coupled nuclear spins.

For the nuclear coherence measurements, we implement spin-to-charge conversion for the readout using two fiber-coupled acousto-optic modulators (AOMs) from AA Opto-Electronic MT250-IR6-Fio-SMO in series for the resonant laser readout for high extinction of the light. The same 705 nm and 1151 nm ionization lasers and pulse scheme were used as in Ref. [8]. All reported error values are 1 SE from the mean.

### *2. Isotopically Purified Diode*

The sample consists of epitaxial 4H-SiC grown by chemical vapor deposition on a 4° off-axis n-type 4H-SiC. The p-type layer thickness is 2 μm and uses naturally abundant Si and C precursor gasses. The i-type layer thickness is ~90 μm and uses isotopically purified Si and C precursor gasses with 99.85% $^{28}Si$ and 99.98% $^{12}C$ as confirmed by secondary ion mass spectroscopy. Single defects are created using irradiation of 2 MeV-electrons at $3\times10^{13}$ /cm$^2$ at room temperature. Single, isolated $VV^0$ formation occurs after annealing at 850 °C in Ar gas for 1.5 hours. Once $VV^0$ defects are formed, we lithographically pattern Ohmic contacts on the top (p-type) and back (n-type) sides of the SiC sample with Ti(30nm)/Al(100)/Ti(10)/Au(30). All contacts were formed by annealing in a quartz tube furnace in Ar at 850 °C for 10 min. We verify the Ohmic contacts through I-V measurements at room temperature and at 5 K.

### *3. Computational Details*

The calculations of the nuclear spin coherence times were done following the recipe described in Ref. [54]. We use the cluster correlation expansion (CCE) method of fourth order [67] and converge the results with the number of clusters included in the expansion. For the results with the electron spin in the $m_s = 0$ state we use generalized cluster correlation expansion [68]. Calculations were done using the pyCCE package [69]. The calculations of the hyperfine interactions were done using DFT with the PBE functional for the supercell with 1438 atoms with the Quantum Espresso package [70]. For the nuclear spins outside of the computed DFT supercell, only the dipole-dipole part of the hyperfine interaction was computed, using a spin density distribution within the DFT supercell.

## APPENDIX B: Optical Stark Shifts

The optical Stark shifts are measured by fitting a Lorentzian on the spin-conserving optically cycling transitions as seen in the photoluminescence excitation scans as in Fig. 5. The grey region that is a guide to the eye for the depletion voltage qualitatively shows similar transitions. The hysteresis between measurements as the diode settles and the low sampling density in the plots shows the voltages not lining up completely the same. Generally, we observe that all three panels show significant drops after -50 V, exactly where the Stark shift begins in Fig. 1(b) to within about one point spacing (10 V).

## APPENDIX C: Nuclear Gates

### *1. Dynamical Decoupling via XY8*

We target a single nuclear spin through nuclear spin resonance spectroscopy using the $VV^0$ spin. The XY8 dynamical decoupling sequence is used, which preserves the coherence of the $VV^0$ spin up to at least 5 s as in Ref. [8], and can be used to perform the nuclear spin spectroscopy. The $VV^0$ spin is optically prepared in the $m_s = 0$ state, followed by a microwave (MW) $\pi/2$ pulse to rotate the spin into its superposition state of $(|0\rangle + |-1\rangle)/\sqrt{2}$. After, a series of $\pi$ pulses in the x (0 degree phase) and y (90 degree phase) quadrature are applied to repeatedly flip the $VV^0$ spin around the two axes. A final +/- $\pi/2$ MW pulse is applied to the $VV^0$ spin to rotate onto either axial pole of the Bloch sphere before optical readout from the $m_s = 0$ transition.

### *2. Initialization and Readout*

The initialization sequence consists of $\pi/2$ rotations of the divacancy electron spin and XY8 sequences on the electron spin that act as conditional rotations of the nuclear spin. The readout gate of the nuclear spin consists of the same sequence for initialization of the nuclear spin, but done in reverse to exchange the spin information from the nuclear spin to the $VV^0$ spin. For these gates of the nuclear spin, we find $\tau = 3.715$ μs will initialize the nuclear spin to $|\uparrow\rangle_{nuc}$ ($|\downarrow\rangle_{nuc}$) when the $VV^0$ is set to $|-1\rangle_{VV0}$ ($|0\rangle_{VV0}$) at the start of the experiment.

### *3. Rabi with Initialization/Readout.*

By fixing the interpulse delay to $\tau' = 3.73$ μs (from original XY8 spectroscopy) and using the nuclear initialization/readout gates with $\tau = 3.715$ μs, the implementation of an XY8 sequence that varies the number of pulses N in the sequence will rotate the nuclear spin about the x-axis of the Bloch sphere.

### *4. Nuclear Ramsey Interferometry.*

A nuclear $\pi/2$ rotation can be performed with N = 34. Depending on the spin state of the $VV^0$ ($m_s = 0$ or -1) during the nuclear free precession, the precession frequency follows

$$f_{\uparrow,\downarrow}^{Ramsey} = \sqrt{(f_L \pm A_{\parallel}/2)^2 + (A_{\perp}/2)^2}$$

The visible difference in the precession frequency confirms that we are indeed addressing a single nucleus and this allows us to extract the hyperfine coupling constants (Fig. 19). From these parameters and the resonance dips of the dynamical decoupling sequence (Fig. 3(a)), we calculate the gyromagnetic ratio to be -5.319 MHz/T, confirming that this coupled nuclear spin is $^{29}Si$ and not $^{13}C$.

We fit the oscillation of each nuclear Ramsey scan at each time point, take the amplitude and associated error, and plot the resulting decay of this amplitude for all cases.

### *5. Hahn Echo.*

We measure the Hahn echo values of the nuclear spin when the electron spin is prepared in either the 0 or -1 state by adding a nuclear $\pi$ sequence between the two nuclear $\pi/2$ rotations.

We verify that none of these values are limited by the $T_1$ decay of the $VV^0$ spin up to 45 s (Fig. 19).

## APPENDIX D: *$^{29}$Si frozen core Hahn echo $\chi^2$ goodness of fit test*

As little decay is observed in the nuclear Hahn echo measurement with $VV^0$ in the $m_s = -1$ spin state, standard least squares regression fitting becomes difficult. Instead, we use a $\chi^2$ goodness of fit test to estimate a lower bound for the nuclear spin Hahn echo decay. We use an exponential stretch factor of n = 2 to n = 3 because these are values normally expected for spins based on their dynamics as seen in Ref. [44] and from our CCE simulations. We use the same statistical analysis as in Ref. [8]. The only difference is that the k value is 6 data points, which corresponds to a degree of freedom of 4. Using these new parameters, we also perform the right-handed test with a confidence level of 99% ($\alpha$ of 0.01) that gives a critical $\chi^2$ value of 13.277. We find that all $\chi^2$ distributions generated for $T_{2,test} < 105$ s exceed this critical value for n = 2. Thus, we place a lower bound estimate of 105 s on the $T_2$ time with 99% confidence. Following the same statistics, we find a $\chi^2_{test}$ lower bound for varying stretch factor from n = 2 to n = 3 and find that all values reach minutes-long $T_2$ times with 99% confidence, specifically 105 s for n = 2 to 85 s for n = 3. The frozen core ensures that there is very limited or no polarization transfer between nuclear spins in the vicinity of the divacancy in which the target nuclear spin is located. We can expect slow noise during the nuclear Hahn echo experiment, so a quasistatic regime is the most reasonable assumption. Thus n = 2 up to n = 3 are more appropriate selections for the stretch factor. Similar coherence measurements of nuclear spins in diamond using the frozen core effect induced by an NV center typically yield exponents n~2 [43, 50]. However, a stronger claim about the stretch factor would require further knowledge of the bath dynamics. From the CCE simulations, the observed number of coupled nuclear spins and hyperfine is not an outlier, but directly consistent with the average expected for any electron spin in this sample (Fig. 23 and 24).

## APPENDIX E: *Microscopic origins of dephasing and relaxation of the electron and nuclear spins*

From the fits of the Hahn echo decays of the electron spin, we find an exponential-decay stretch factor of $n \approx 1.5$ in the undepleted state, corresponding to dynamic interaction of the $VV^0$ electron spin with a paramagnetic spin bath [55, 71]. In depletion, the stretch factor converges to $n \approx 2$, suggesting that the defect is limited by 1/f-type noise [72], which can be explained by coupling to an interacting dipolar spin bath [73] as in Fig. 10. We also find the stretch factor of the non-frozen core Hahn-echo decay to be $n = 2$ for the nuclear spin, confirming the 1/f noise seen by the $VV^0$. Here we discuss possible origins of relaxation and dephasing processes of the electron and nuclear spins. The first candidate consists of dipole-coupled paramagnetic impurities, which are known to be a main source of decoherence in solid-state defect-spin systems. We briefly summarize features of such interacting spin baths; for more details, see, e.g., [55, 73]. The dephasing noise induced by such spin baths can be captured by a characteristic dipolar coupling rate: $R \propto n_B$, where $n_B$ denotes the bath spin density. In the regime where the bath spins are sufficiently sparse, it induces a Gaussian decay in electron or nuclear spin coherence function as measured in a Ramsey sequence, where the coherence time has a simple scaling relation with the bath spin relaxation time $T_{1,B}$:

$$T_2^* \propto \sqrt{T_{1,B}}$$

The spin bath also introduces a spin relaxation process $T_1$ on the measured electron or nuclear spin. When the paramagnetic impurities are highly off-resonant with the measured spin, the relaxation time scales as $T_1 \propto T_{1,B}$. Coincidentally, the enhancement in the nuclear spin relaxation time and its Ramsey dephasing time (with $m_s = 0$) exhibit the same scaling as that induced by a sparse dipole-dipole coupled spin bath. However, we are not able to identify or measure such paramagnetic impurities in the current experiment, and more investigations are needed to verify this scaling.

Another possible source of decoherence is laser-induced ionization of the electron spin, which in turn leads to nuclear spin relaxation and dephasing. In this case, given the laser-induced linewidth of the electron spin $\gamma_{opt}$, the optically-induced nuclear spin relaxation rate is given by

$$1/T_{1,opt} = A_\perp^2 / (16\, \gamma_{opt})$$

where $A_\perp$ is the transversal hyperfine coupling strength. Making use of parameters $A_\perp = 9.4$ kHz and $\gamma_{opt} \sim 20$ MHz as extracted from measurements, the optically-induced nuclear spin lifetime during laser illumination is $T_{1,opt} \simeq 4$ s. Combining this estimate with the 40 μs laser pulse duration during a 100-ms-long nuclear spin Ramsey sequence, we compute a laser contribution to an average nuclear spin $T_1$ to be around 2.5 hours, which is of the same order of magnitude as the observed quantity. However, this mechanism would predict a shorter nuclear spin $T_1$ with a decreasing electron spin optical linewidth, which contradicts the measured trend. We thus do not expect the optically-induced linewidth to be a dominant mechanism for nuclear spin relaxation.

## APPENDIX F:. *Photon Counting Statistics and Readout Fidelity of Nuclear Spin*

### *1. Hidden-Markovian model for determining the threshold and $T_{1n}$.*

To determine the readout fidelity and the $T_{1n}$, we first find a threshold for the electron Ramsey scans from our 22.5 hr scans at depleted and non-depleted biases (Fig. 2(b)). These measurements are affected by the telegraph noise caused by the weakly-coupled $^{29}Si$ switching from its spin up to spin down state. We can use a model that encompasses the correlated nature of nuclear spin fluctuations such as the Hidden Markovian Model where hidden states (the nuclear up and down spin states) probabilistically switch over time and affect the noisy, repetitive readout from the Ramsey measurement. We select a region of interest from 104 μs - 144 μs for -80 V (Fig. 2) and 104 μs - 120 μs for 0 V (Fig. 2) and do a rolling average window of size 10 and 5, respectively. From the thresholding and processing, we find the distribution of times for nuclear spin state up and down and the respective $T_{1n}$ during this repetitive readout of the nuclear spin.

### *2. Readout Fidelity of nuclear repetitive readout.*

We take the definition of fidelity following Ref. [8]:

$$F = 1 - ½(1 - p_{0|0} + p_{0|1}) = ½(p_{0|0} + p_{1|1}) \qquad (1)$$

where $p_{0|0}$ is the normalized true positive probability, $p_{0|1}$ is the normalized false positive probability, $p_{1|1}$ is the normalized true negative probability and $(1 - p_{0|1}) = p_{1|1}$. We can determine the single-shot cutoff of n photons from Eq. (S1) and calculate the single shot nuclear down fidelity as:

$$F_{nuclear} = \frac{1}{2} \left( \Sigma_0^n p_\uparrow / \Sigma_0^\infty p_\uparrow + \Sigma_0^n p_\downarrow / \Sigma_0^\infty p_\downarrow \right) \quad (2)$$

where $p_\uparrow$ and $p_\downarrow$ are the unnormalized photon distributions for the up and down nuclear states, respectively. Eq. (2) is used to calculate the data in Fig. 2. The cutoff n is chosen to maximize F.

### **APPENDIX G:** ***22.5 Hour Ramsey Processing***

From the data in Fig. 2, we take the FFT of the data to show the drifts in the $VV^0$ spin frequency of both 0 V and 80 V of reverse bias in the diode (Fig. 11 and 12). We also plot the frequency changes and discrete jumps as a function of the time (Fig. 13 and 14) to further study the noise behavior during the 22.5 hour scans and find a mean and standard deviation of the frequency binning from 0 V and 80 V of reverse bias. We observe that the variance in the frequencies is 40% less in the 80 V reverse bias scan when compared to the 0 V scan (Fig. 15 and 16). Furthermore, while fitting the $T_2^*$ as a function of the amount of sweeps (assuming a fixed n = 2) and monitoring the change in the Ramsey decay, we find that the values merge after 1.5 hours (Fig. 17). The variation and changes in the magnetic frequency affect these times and the saturation of the $T_2^*$ time at around 100 μs.

### **APPENDIX H:** ***Depletion of paramagnetic impurities***

We can deplete the p-i-n device of all the $VV^0$ defects in the intrinsic region of the junction by applying a larger bias as demonstrated in Ref. [24]. A large reverse bias can simultaneously deplete all $VV^0$ in the device's intrinsic region. We suggest that the origin of the magnetic noise reduction discussed in the main text is from the depletion of certain types of paramagnetic species based on previous ab-initio results (i.e. supported by theoretical evidence). Existing ab-initio DFT results position the charge transition levels in the bandgap [58] and EPR experiments show the associated spin multiplicity of the most common impurities in SiC [60]. In the absence of bias, the $V_C^-$ (or $V_C^0$) charge state has spin ½ (or spin 0), and $N^0$ is in the charge state with S = 1/2 [74]. Under bias, photoexcitation removes charges from N, which drift away, leaving $N^+$ with S = 0, which is never recharged. On the other hand, $V_C$ starts in the negative charge state with S = 1/2. Under 1.75 eV (705 nm) excitation and diode depletion $V_C^-$ (S = 1/2) converts to $V_C^0$ (S = 0), which is likely to be stable under 705 nm illumination or cycle between + and 0 charge state [59]. Fig. 27 illustrates this argument more clearly with appropriate energy and charge states of the traps within the band gap of the material. The Ramsey coherence time is significantly longer than the ab-initio cluster correlated expansion (CCE) predicted value of 54 μs at the isotopic concentration of the SiC [12]. On the other hand, even with the coherence improvement, the measured Hahn echo value is an order of magnitude lower than predictions from ab initio CCE calculations, requiring further investigation. The claim of reduction in paramagnetic noise is therefore not directly verified, but the noise reduction is proven to be magnetically-related and diode dependent from the data, and photodynamic understanding from ab-initio calculations support the conversion to spin-free defects of the major impurities.

### **APPENDIX I:** ***Noise Analysis of the Electron Hahn and Ramsey***

Assuming that the noise is stationary and Gaussian, we can extract certain ranges of the bath noise spectral density from the electron Ramsey and Hahn echo measurements. Denote by B(t) the bath noise operator coupled to the electronic spin Pauli operator $\sigma_Z$ and define $S(t) \coloneqq \langle B[0],B(t)\rangle/2$ as the noise autocorrelation function. The electronic coherence (i.e. the off-diagonal element of the central spin density matrix in Pauli $\sigma_Z$ eigenbasis) is related to a coherence decay factor ξ(t) as $\langle \sigma_-(t) \rangle = \exp(-\xi(t))$. The decay function ξ(t) depends on the modulation function f(t) of the electronic spin coupling to its environment, and the noise autocorrelation S(t) as:

$$\xi(t) = \frac{1}{2} \int_0^t dt_1 \int_0^t dt_2 \, S(t_1 - t_2) \, f(t_1) \, f(t_2)$$

We can rewrite the equation above in terms of the power spectral density (PSD) of the noise (also known as noise spectral density (NSD))

$$S[\omega] = \int_0^{+\infty} e^{i\omega t} S(t) \, dt$$

and the Fourier transform of the modulation function $f[\omega] = \int_0^t e^{i\omega t_1} f(t_1) \, dt_1$, as

$$\xi(t) = (1/4\pi) \int_0^\infty S[\omega] \, |f[\omega]|^2 \, d\omega = (1/4\pi) \int_0^\infty S[\omega] \, F[\omega] \, d\omega \quad (S3)$$

The filter function $F[\omega] \coloneqq |f[\omega]|^2$ describes the spectral response of the central spin to bath noise. For Ramsey, the filter function equals

$$F_{Ramsey}[\omega] = 4 \sin^2(\omega t/2) / \omega^2$$

For Hahn echo, the filter function equals

$$F_{Hahn}[\omega] = 16 \sin^4(\omega t/4) / \omega^2$$

We can use experimentally measured coherence function decay factor, ξ(t), to infer the noise spectral density. However, this reconstruction is only reliable in regimes of frequency ranges that are accessible in measurement data. For Ramsey and Hahn echo measurements, this typically means that we can only reliably reconstruct the low-frequency regime of the bath noise spectrum.

For quasistatic noise, whose noise spectrum only has nonzero values at frequencies lower than the resolution of the experiment, we can approximate the filter function as $\omega t \ll 1$:

$$F_{\text{Ramsey}}[\omega] \approx t^2$$

In this regime, the central spin coherence decays as a stretched exponential function with stretched exponent 2. Conversely, if the experimentally measured coherence function's decay stretched exponent is close to 2, then the inversion problem defined by Eq. (S3) becomes ill conditioned, and we can only extract an integrated noise spectral strength from the data.

Fig. 28 depicts the reconstructed noise spectrum for both voltages. To extract the noise spectral density, we first extract envelope functions for all experimentally measured coherence data to eliminate phase oscillations due to strongly coupled bath spins. We then use the logarithm of the envelope function as input data for $\xi(t)$ and find the best numerical fits that satisfy Eq. (S3). The resulting noise spectra are shown in Fig. 28. For both voltages, the Ramsey coherence data reveals strong low-frequency noise, but the Ramsey measurement does not enable sufficient frequency resolution in the reconstructed noise spectrum. The extracted noise spectrum in the depleted regime is significantly smaller than the undepleted regime, demonstrating the efficacy of voltage for tuning the electronic bath noise. Figures S25 and S26 show the coherence functions under the Ramsey sequence as predicted by the reconstructed noise spectrum in Fig. 28. The corresponding experimentally measured values are also included for reference. We see that the extracted noise spectra accurately reproduce the experimentally measured decay envelopes, up to measurement uncertainties.

## APPENDIX J: *Electrically-Limited Ramsey Linewidth*

Assuming the spin transition linewidth experiences the same charge-noise broadening mechanism as the optical lines, we can use the undepleted optical linewidth to predict the Ramsey $T_2^*$ spin linewidth. Charge noise broadens emitter optical lines through spectral diffusion (via the known electric dipole of the optical transition), so the assumption is that charge noise affects the spin linewidth the same way through the known electric dipole of the spin ground state. The measured $VV^0$ optical undepleted dephasing rate compared to the theoretical lifetime limited linewidth multiplied by the electric field susceptibility of the defect therefore gives the expected depleted $T_2^*$ linewidth. From Falk et al. [75], which reports the electric field susceptibility of the PL2 $VV^0$, we find that the expected contribution of electronic noise cannot explain the spin linewidths, therefore there are additional magnetic noise contributions. From these calculations, we find the electrically-limited Ramsey dephasing linewidth to be ~120 Hz. However, our measured value is significantly larger, by over an order of magnitude (3 kHz). Therefore, our results cannot be explained strictly from charge noise.

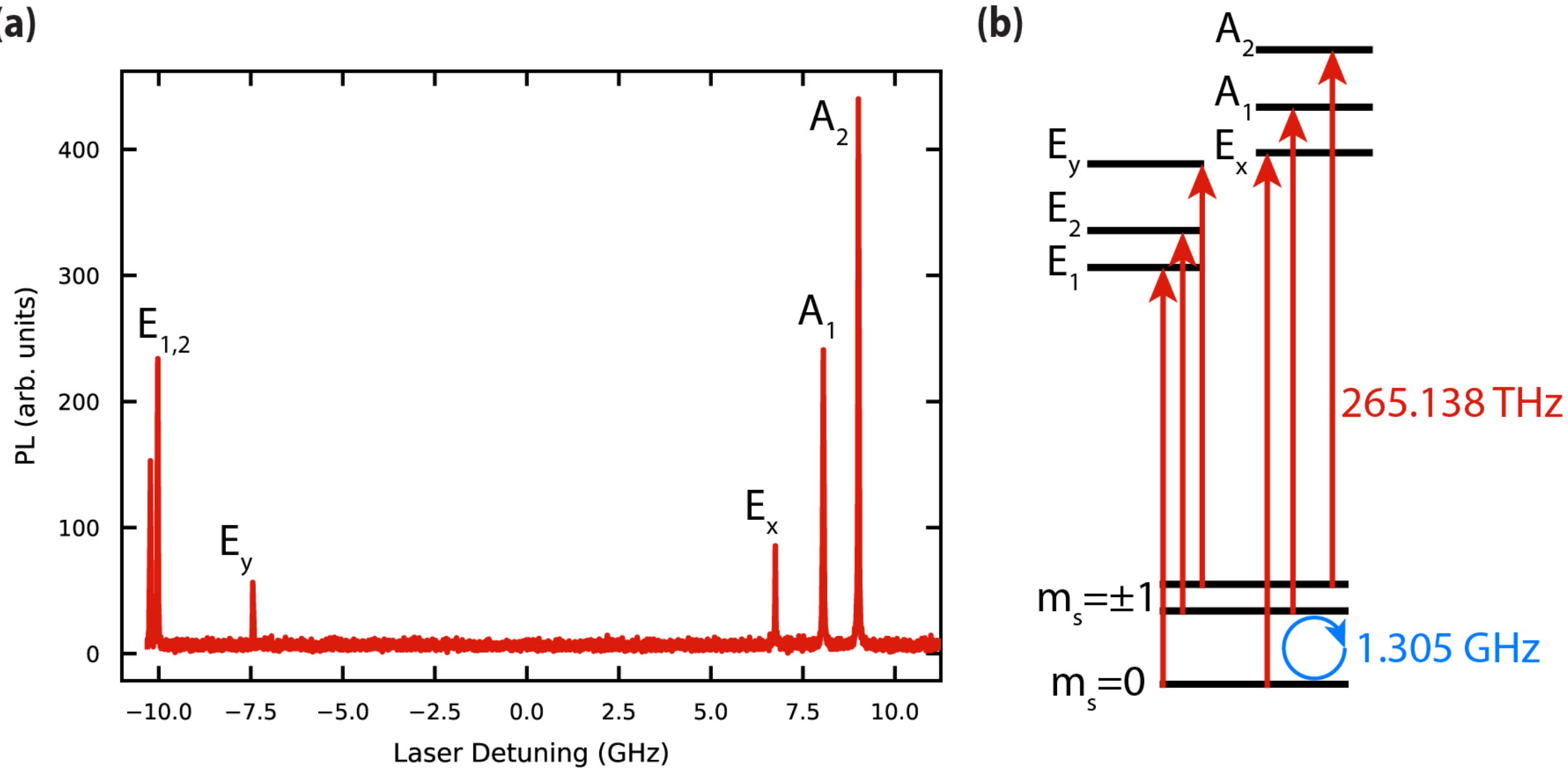

FIG 5. Photoluminescence excitation scan of a PL2 $VV^0$. (a) Photoluminescence excitation (PLE) spectrum of a single PL2 divacancy reveals its six spin-selective optical transitions with a detuning of a center laser frequency of 265.138 THz at T = 5 K with continuous microwave driving of the $m_s = 0 \leftrightarrow m_s = -1$ transition. The $E_x$ and $E_y$ optical transitions preserve $m_s = 0$ character, while the $E_1$, $E_2$, $A_1$, and $A_2$ transitions correspond to the $m_s = \pm 1$. (b) Diagram of spin-conserving optically cycling transitions from a PLE measurement for PL2 divacancies.

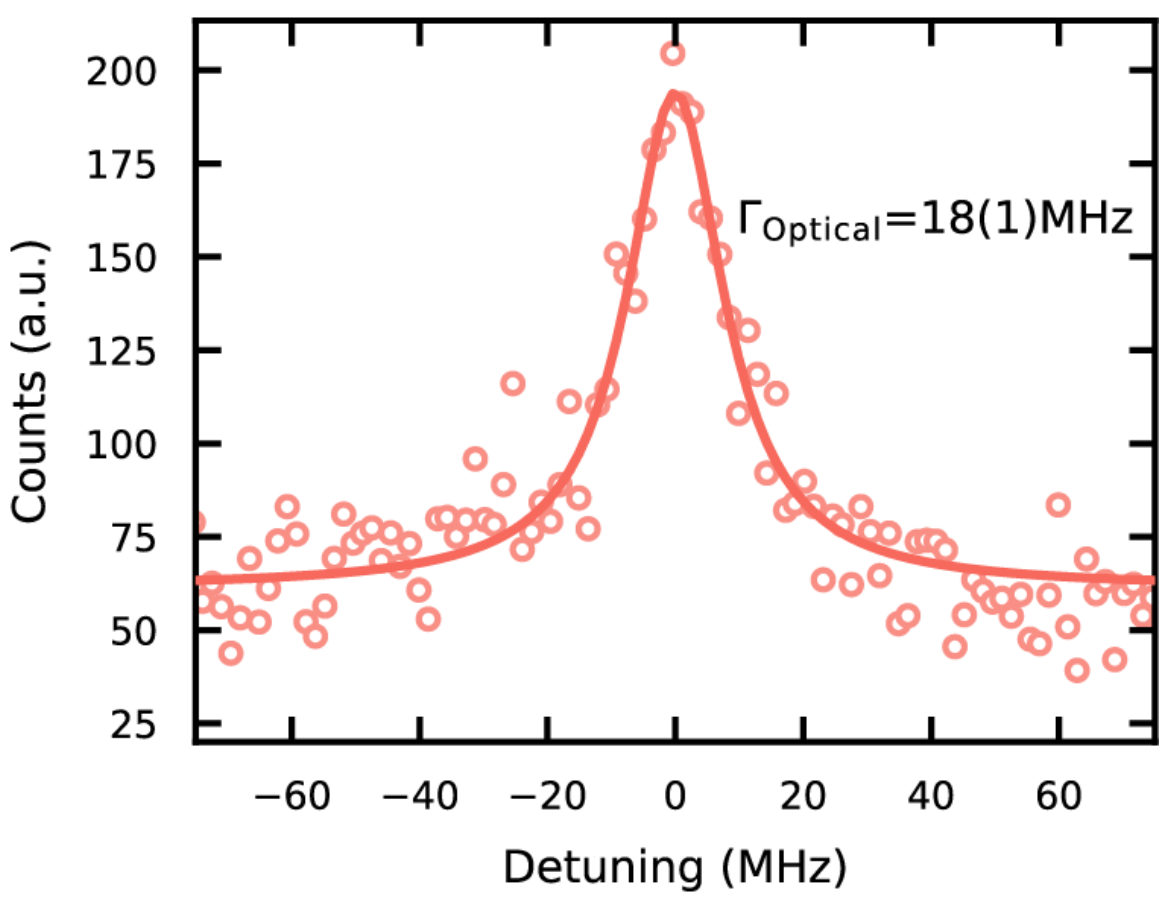

FIG 6. Near Lifetime-Limited Averaged PLE scan of $E_x$ transition. Averaged PLE scan of the $E_x$ transition from four electro-optic modulator sweeps in Fig. 1(c) with a Lorentzian fit.

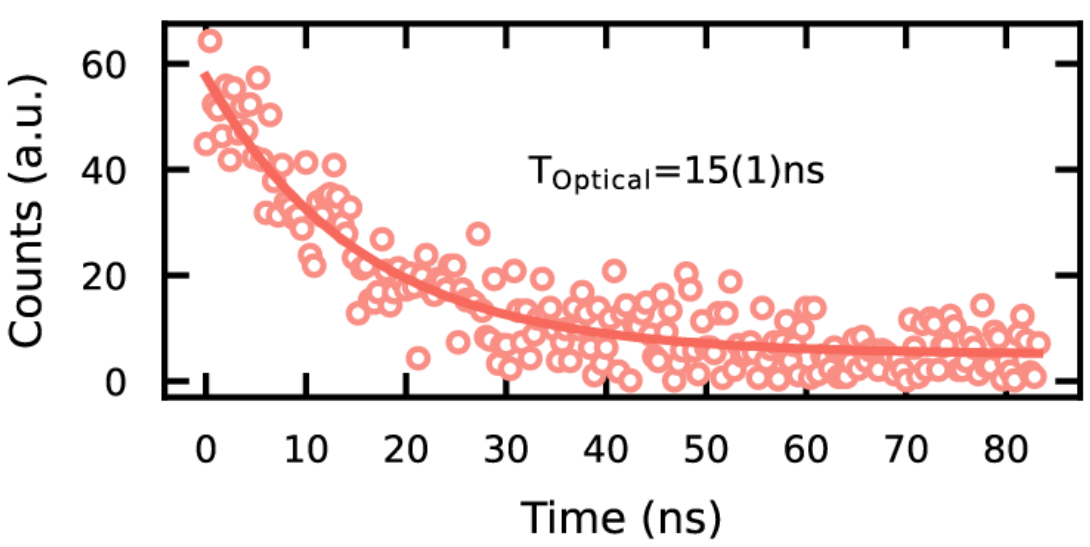

FIG 7. Optical Lifetime of the $E_x$ transition of $VV^0$. Decay of the $VV^0$ emitter lifetime of its $E_x$ transition while under reverse bias of 80 V.

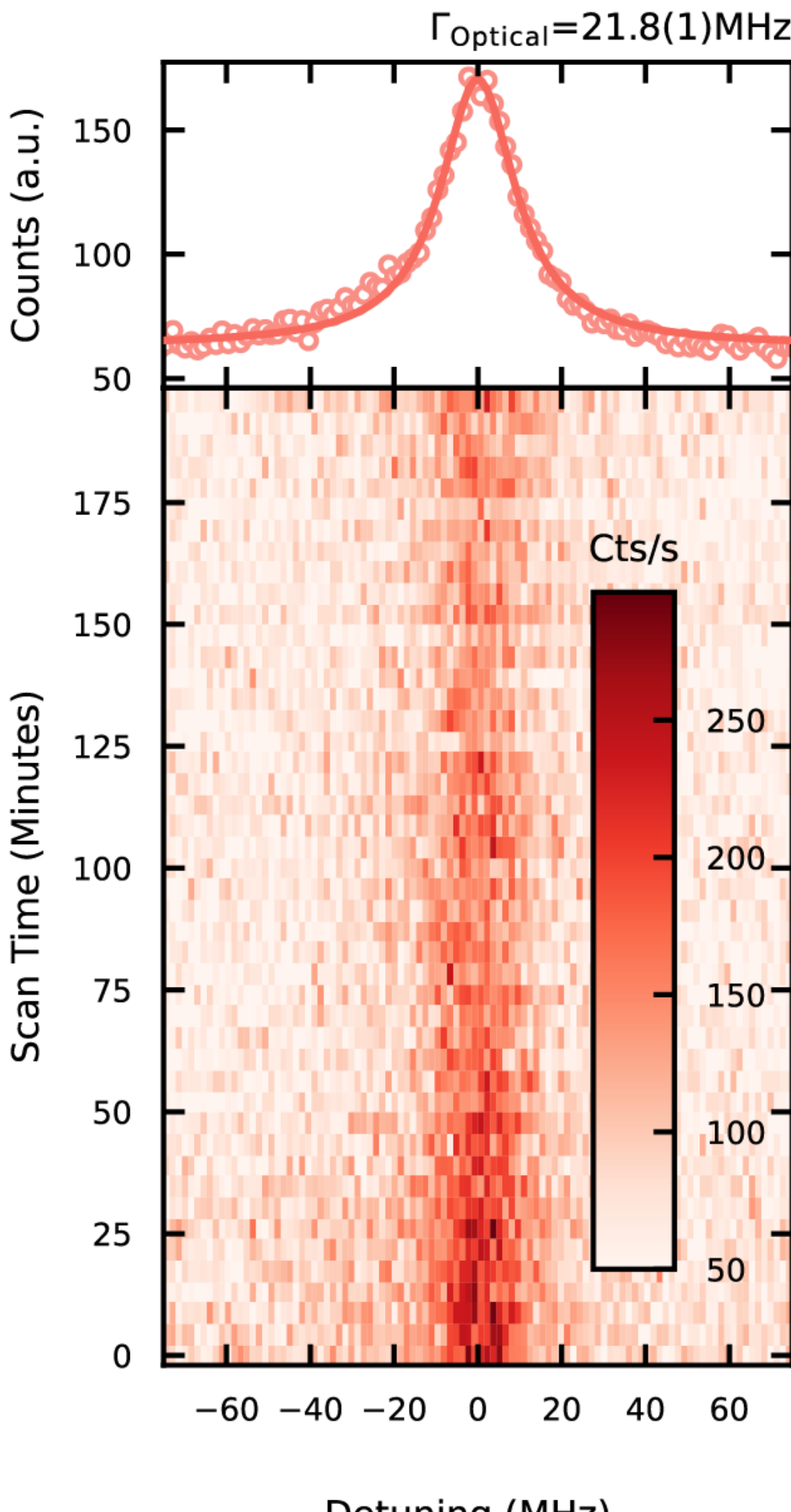

FIG 8. Line Stability of the $E_x$ transition of $VV^0$. (top) Averaged value of a one-hour scan of the $E_x$ transition. (bottom) A three-hour electro-optic modulator sweep of the $E_x$ line of a single PL2 $VV^0$ as a function of time. Each scan on the horizontal axis corresponds to ~4 minutes.

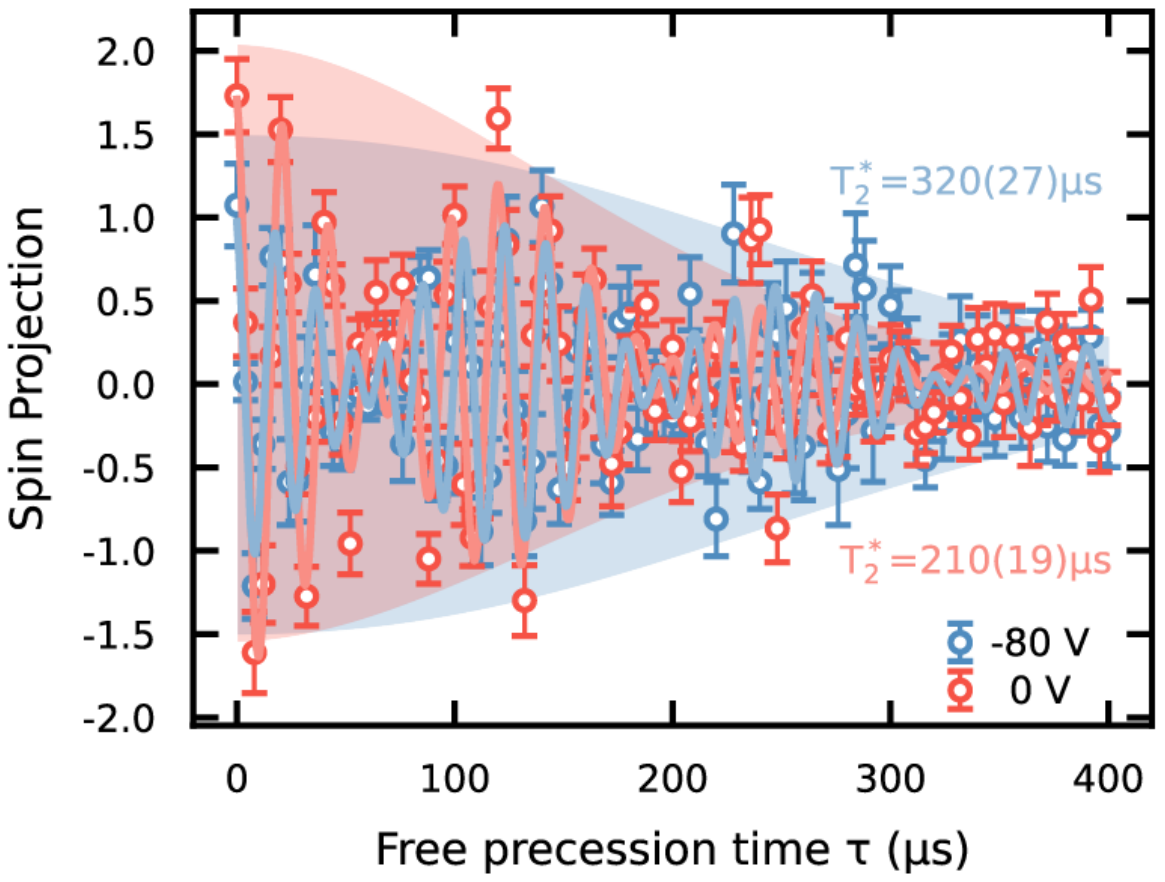

FIG 9. Single Quantum Ramsey. Dephasing time and fits from a $VV^0$ single quantum Ramsey experiment for 80 V of reverse bias and 0 V (red) of bias applied. The different colored shaded areas demonstrate the normalized Gaussian decay envelope of each dataset.

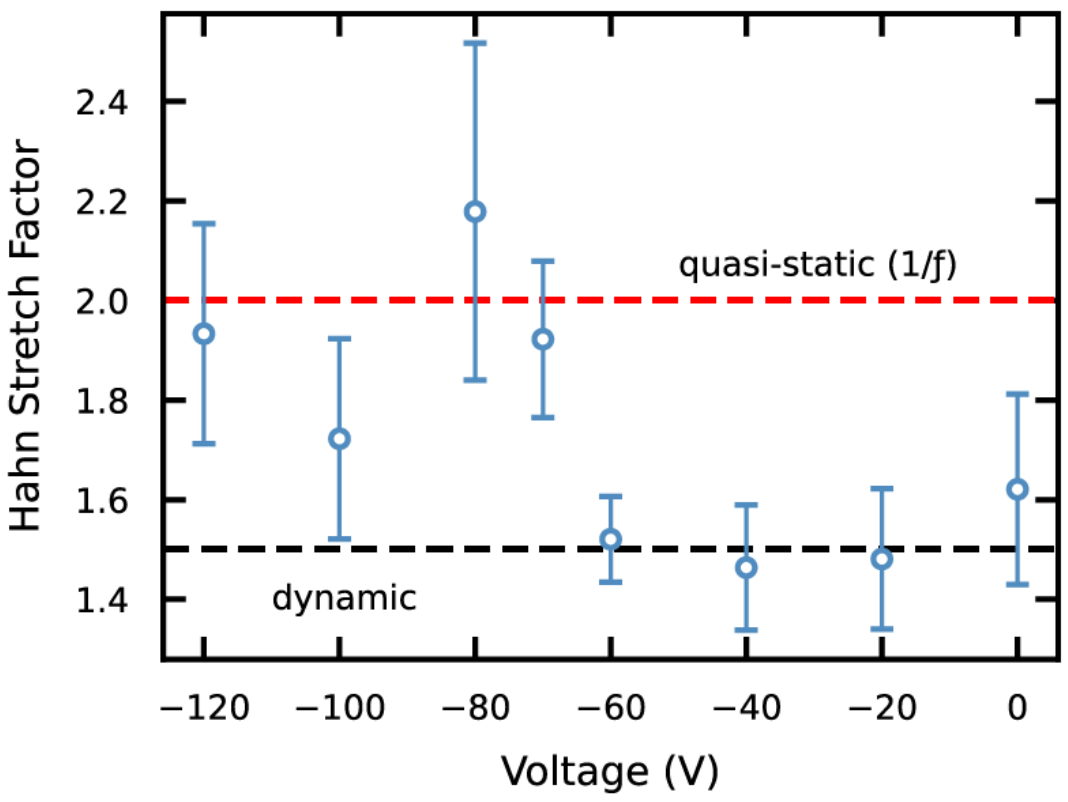

FIG 10. Hahn Echo Decay Stretch Factor versus Voltage. The $VV^0$ Hahn echo decay follows a stretched exponential $\exp(-(\tau/T_2)^n)$ with the stretch factor n. The stretch factor begins around 1.5 in undepleted material and increases to 2 at depleted values. Dashed lines are guides to the eye.

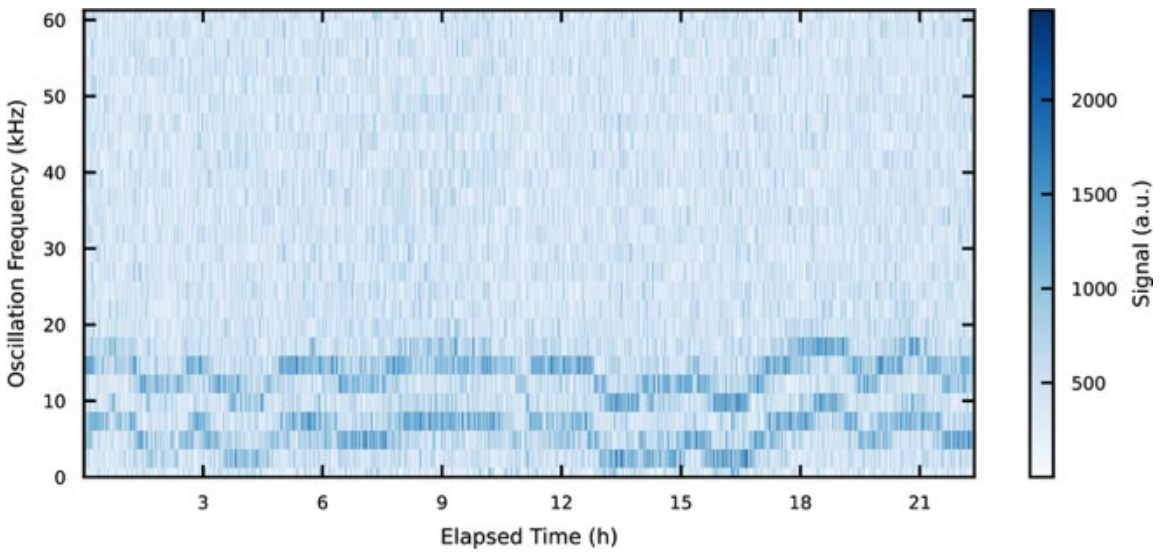

FIG 11. Magnetic Spectrogram of Day-Long Scan of $VV^0$ Ramsey under 0 V of Reverse Bias. The FFT of each $VV^0$ Ramsey experiment throughout the 22.5 hr scan is plotted as a function of the entire experiment time. The two signal peaks are from a strongly coupled nuclear spin causing a splitting of the detuning frequency. A detuning of 10 kHz is used.

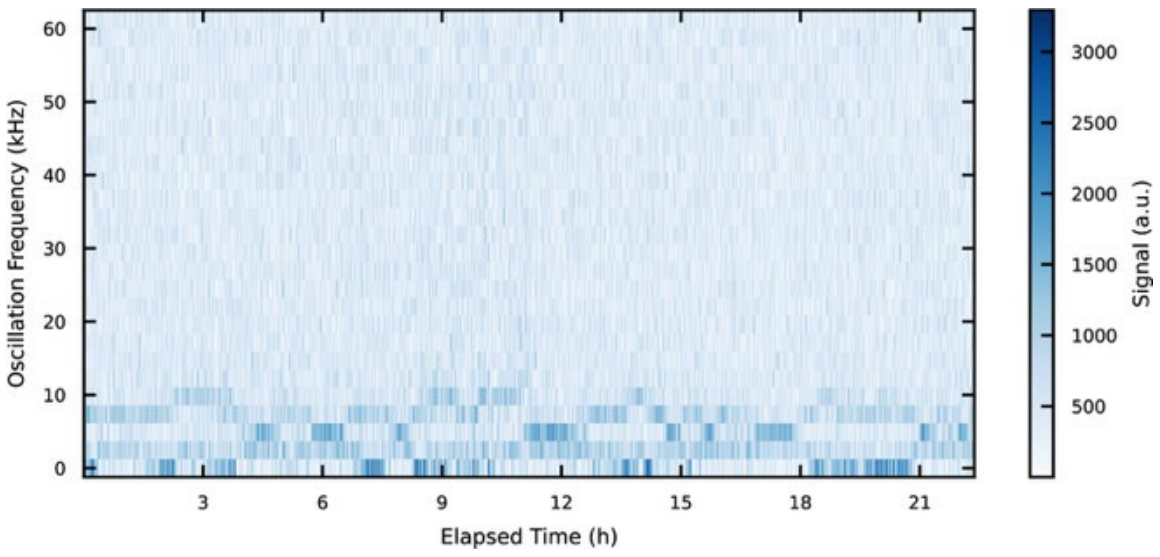

FIG 12. Magnetic Spectrogram of Day-Long Scan of $VV^0$ Ramsey under -80 V of Reverse Bias. The FFT of each $VV^0$ Ramsey experiment throughout the 22.5 hr scan is plotted as a function of the entire experiment time. The two signal peaks are from a strongly coupled nuclear spin causing a splitting of the detuning frequency. A detuning of 10 kHz is used.

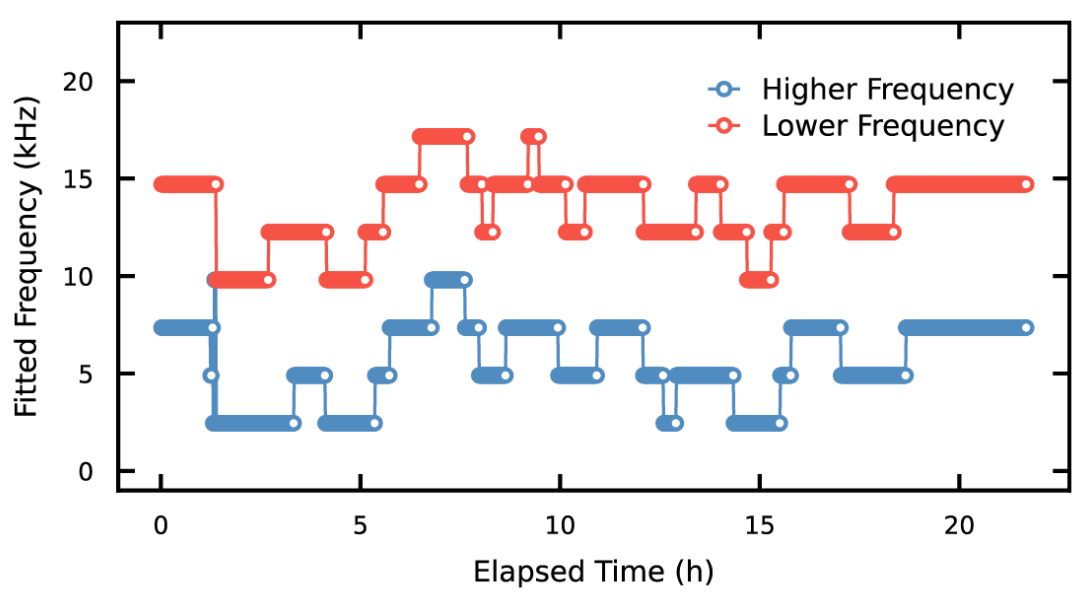

FIG 12. Ramsey frequency jumps during 22.5 h scan at 0 V. Higher and lower frequencies plotted of the FFT of each $VV^0$ Ramsey scan taken during the 22.5 hr scan. There are 8 transitions that are seen total.

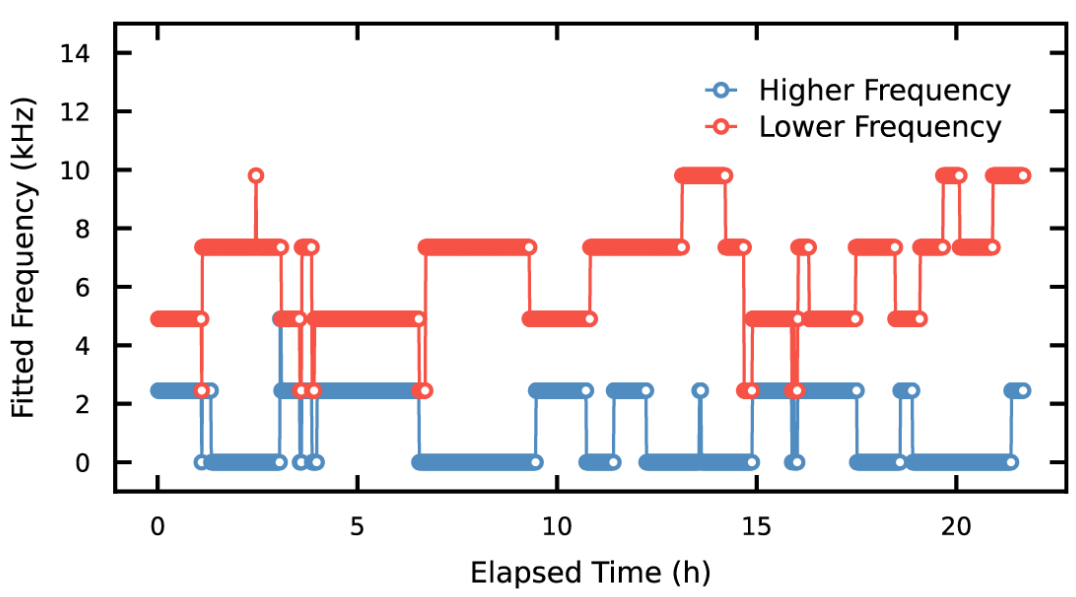

FIG 13. $VV^0$ Ramsey frequency jumps during 22.5 h scan at -80 V. Higher and lower frequencies plotted of the FFT of each $VV^0$ Ramsey scan taken during the 22.5 hr scan. There are 4 transitions that are seen in total.

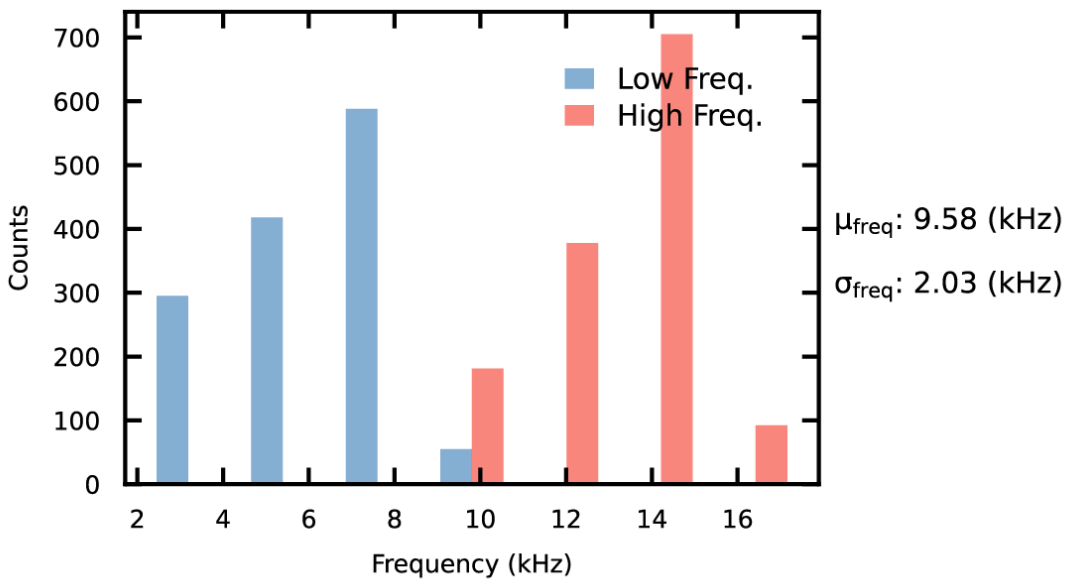

FIG 14. $VV^0$ Ramsey frequency Jump Histograms for 0 V. The binning for low (blue) and high (red) frequencies during the 22.5 hr scan shows a mean of 9.58 kHz with a standard deviation of 2.03 kHz.

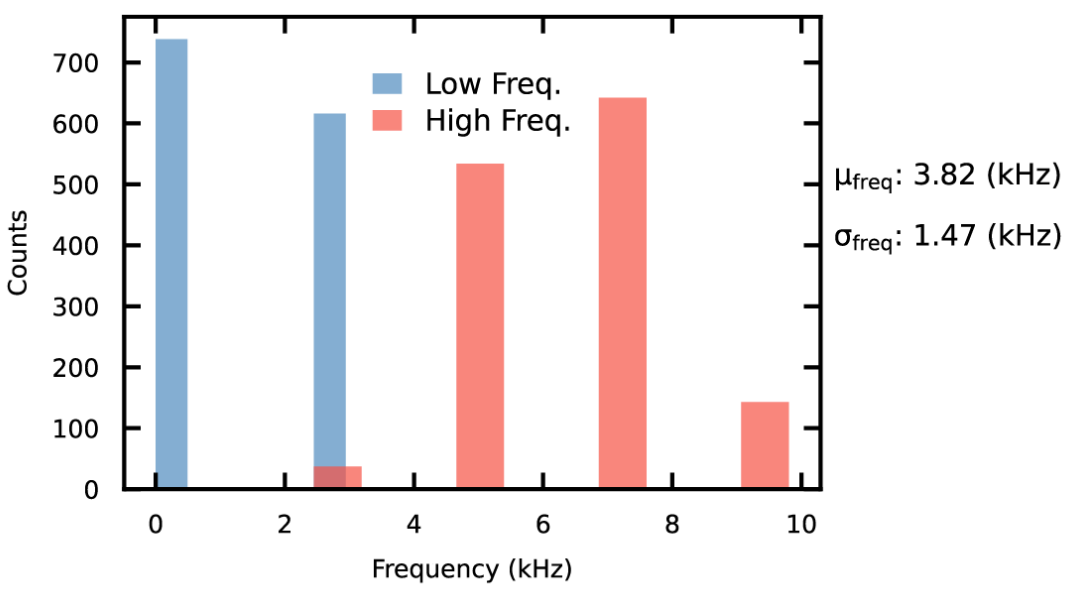

FIG S15. $VV^0$ Ramsey frequency jumps during 22.5 h scan at -80 V. The binning for low (blue) and high (red) frequencies during the 22.5 hr scan shows a mean of 3.82 kHz with a standard deviation of 1.47 kHz.

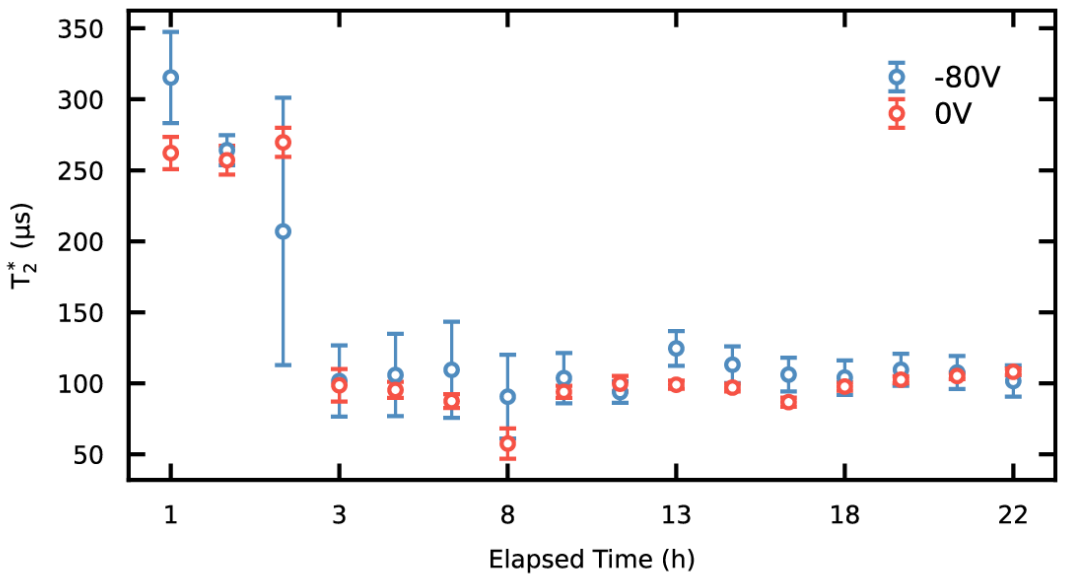

FIG 16. Average $VV^0$ $T_2^*$ versus averaging time. The $VV^0$ $T_2^*$ shows little difference after two hours of averaging, and drop off to a value of about 100 μs afterwards. This likely arises from experimental instabilities during the averaging such as slow magnetic field drifting.

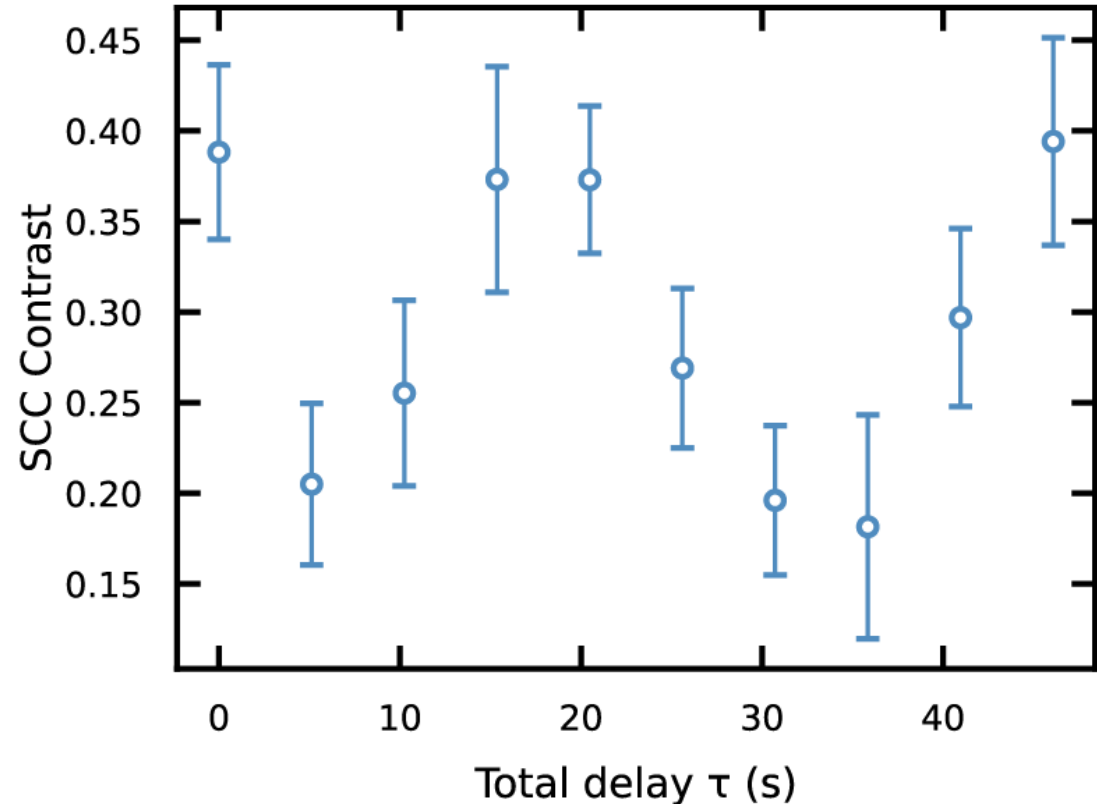

FIG 17. $VV^0$ $T_1$ Spin Relaxation measurement using Spin-to-Charge Conversion. No decay from the $VV^0$ spin $T_1$ measurements confirms that all coherence times are not $T_1$ limited.

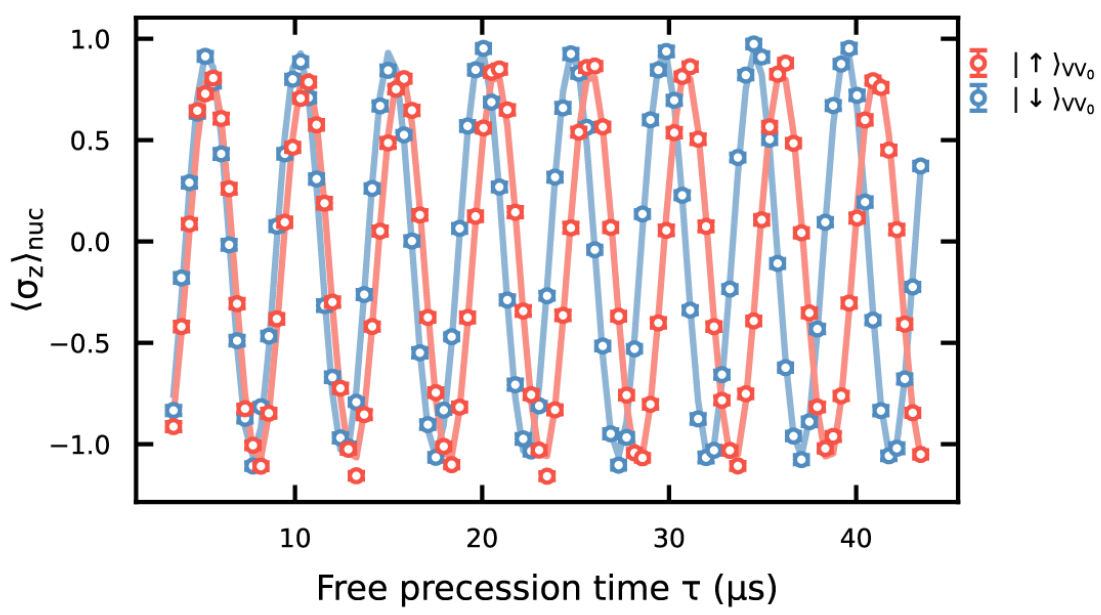

FIG 18. Ramsey Interferometry of single $^{29}$Si. The $VV^0$ electron is initialized in $m_s = 0$ (red) or $m_s = -1$ (blue) after the initial nuclear $\pi/2$ rotation as in Fig. 4(a). The nucleus precesses at different rates depending on the $VV^0$ initial spin state.

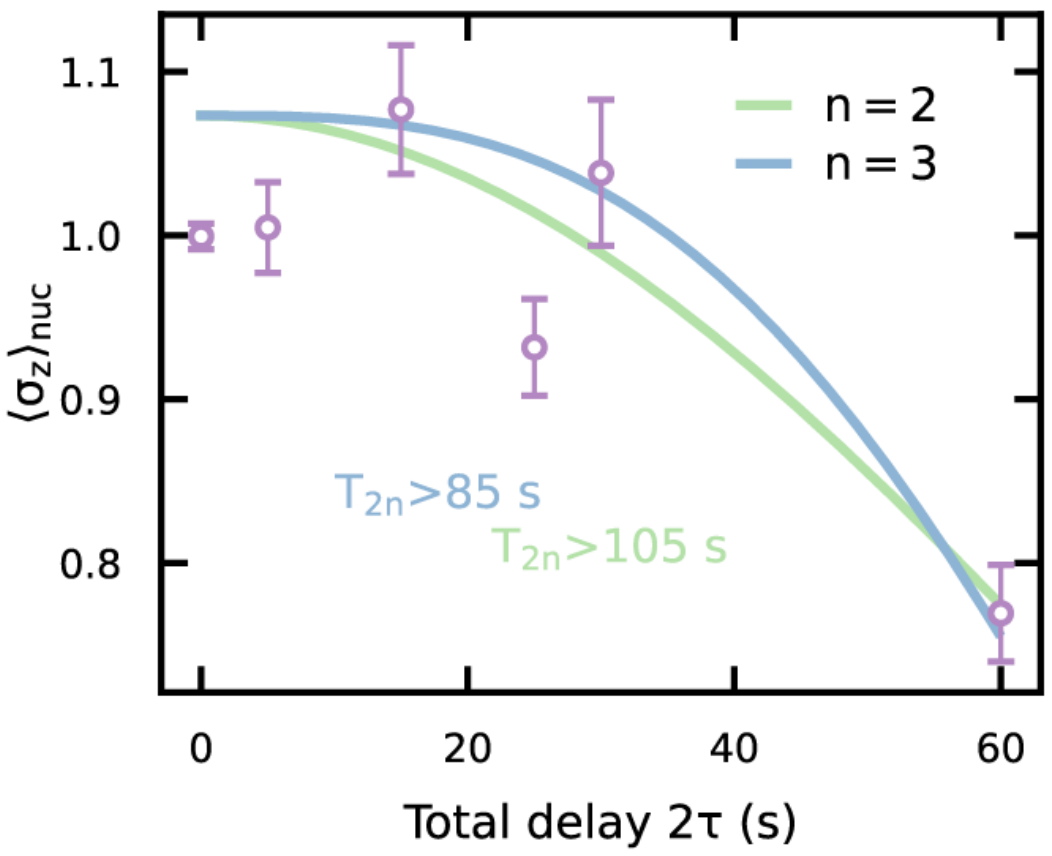

FIG 19. $\chi^2$ lower bound of $T_2$ nuclear spin Hahn echo versus exponent stretch factor. The $^{29}$Si nuclear spin Hahn echo decay fit stretch factor is varied from n = 2

to n = 3 for the $\chi^2$ test. The different outcomes for the $T_2$ lower bound of the test based on the stretch factors are plotted showing minutes-long Hahn echo decays.

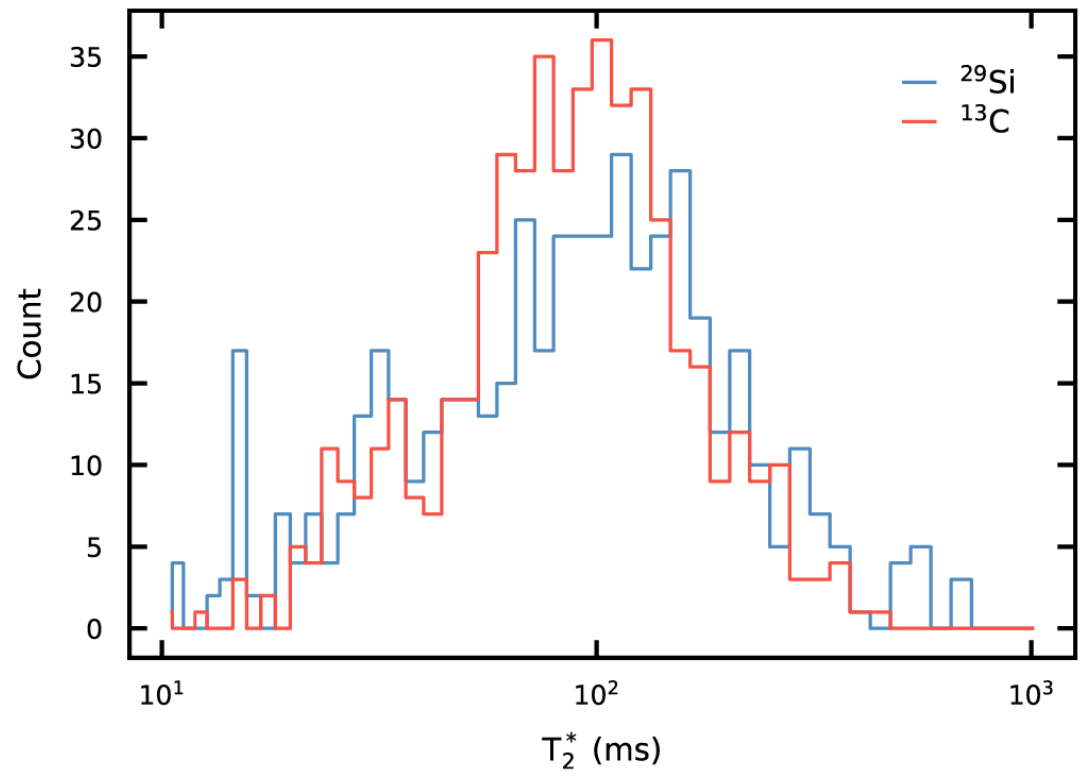

FIG 20. Nuclear $m_s$ = 0 $T_2^*$ distribution. Histograms showing the CCE-calculated Ramsey decay times of $^{13}C$ and $^{29}Si$ when the $VV^0$ is prepared in the $m_s$ = 0 spin state.

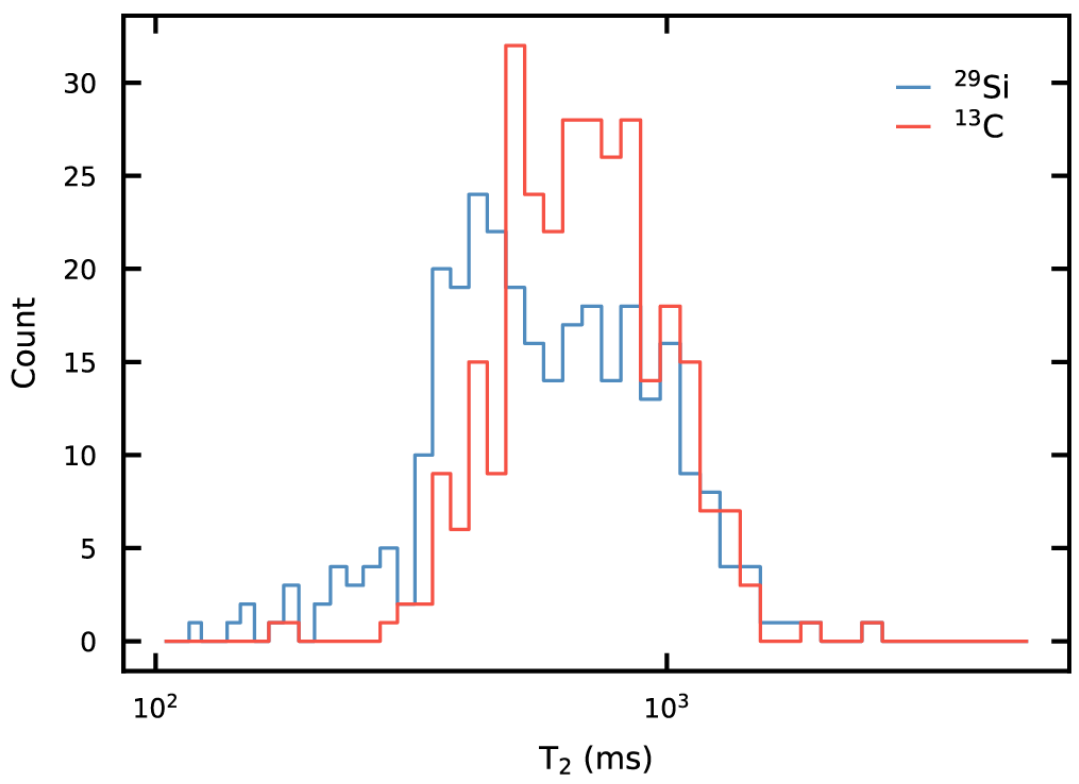

FIG 21. Nuclear $T_2$ Hahn echo distribution. Histograms showing the CCE-calculated Hahn echo decay times of $^{13}C$ and $^{29}Si$ when the $VV^0$ is prepared in the $m_s$ = 0 spin state.

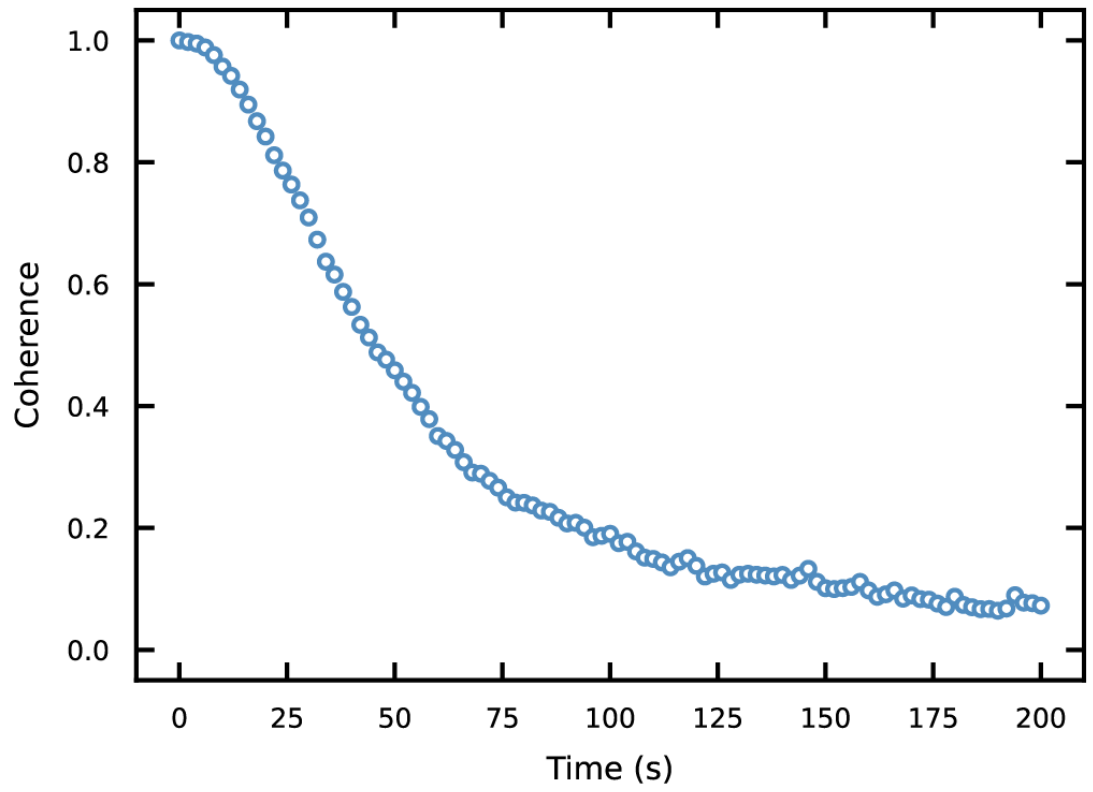

FIG 22. CCE $^{29}Si$ Nuclear $T_2$ decay ($m_s$ = -1). The Hahn echo decay for a candidate $^{29}Si$ nuclear spin with properties $A_{\parallel}$ = 7.9205 kHz and $A_{\perp}$ = 7.8903 kHz when the $VV^0$ is prepared in the $m_s$ = -1 spin state. The approximate distance from the electron spin is 1.385 nm. The $T_2$ time is 59.0(3) s with n = 1.56(2).

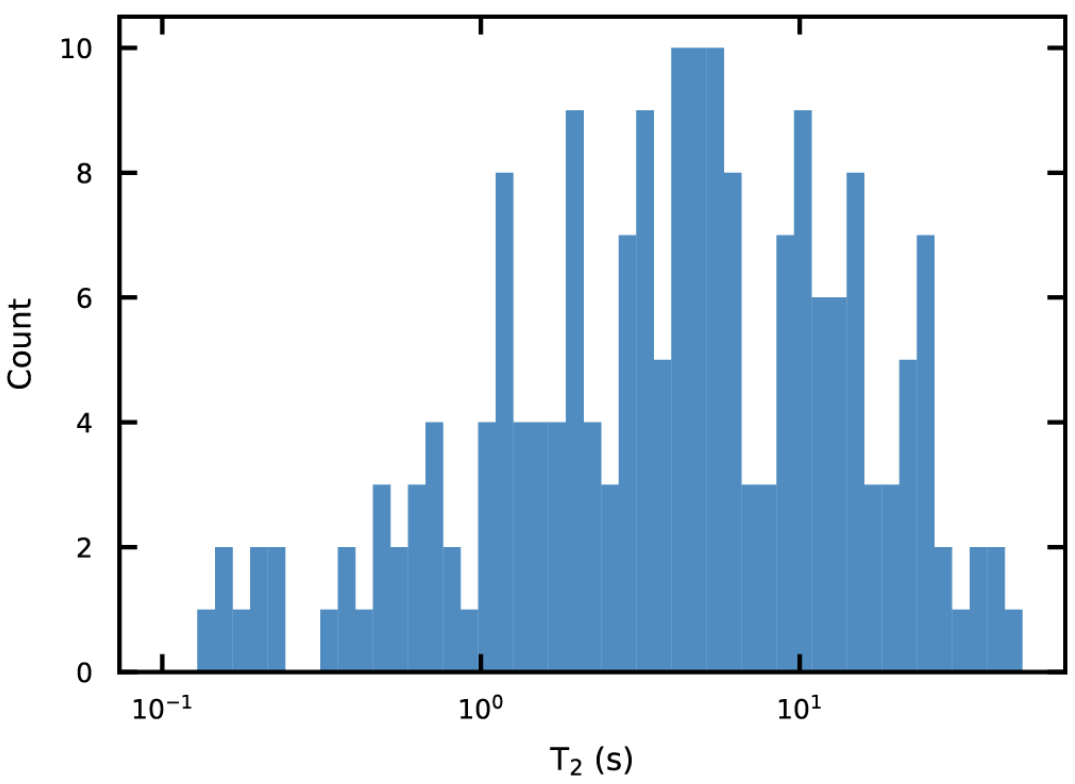

FIG 23. Electron-limited $^{29}Si$ Nuclear $T_2$ Hahn Echo. Distribution of $T_2$ Hahn echo of a $^{29}Si$ nuclear spin at electron density of $10^{15}$ $cm^{-3}$. The coherence times are reported in seconds.

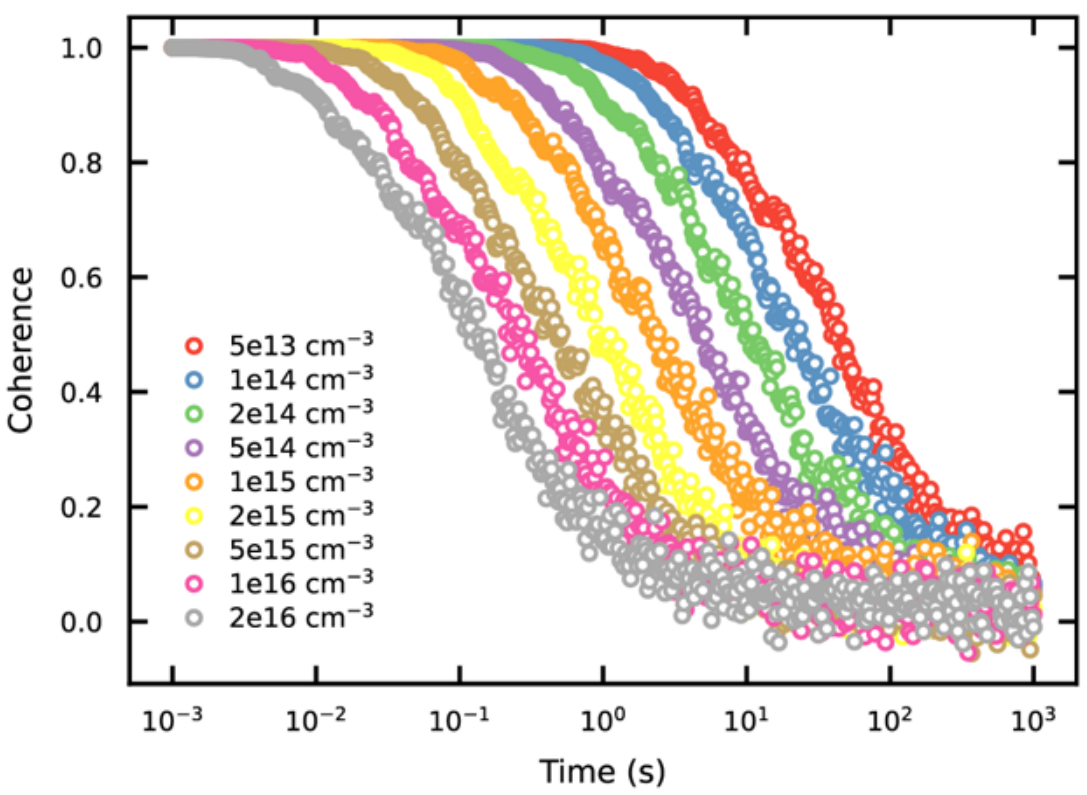

FIG 24. $^{29}Si$ Nuclear Hahn echo $T_2$ vs electron spin density waterfall plot. Average coherence of the $^{29}Si$ nuclear spin as a function of the electron density as calculated with CCE.

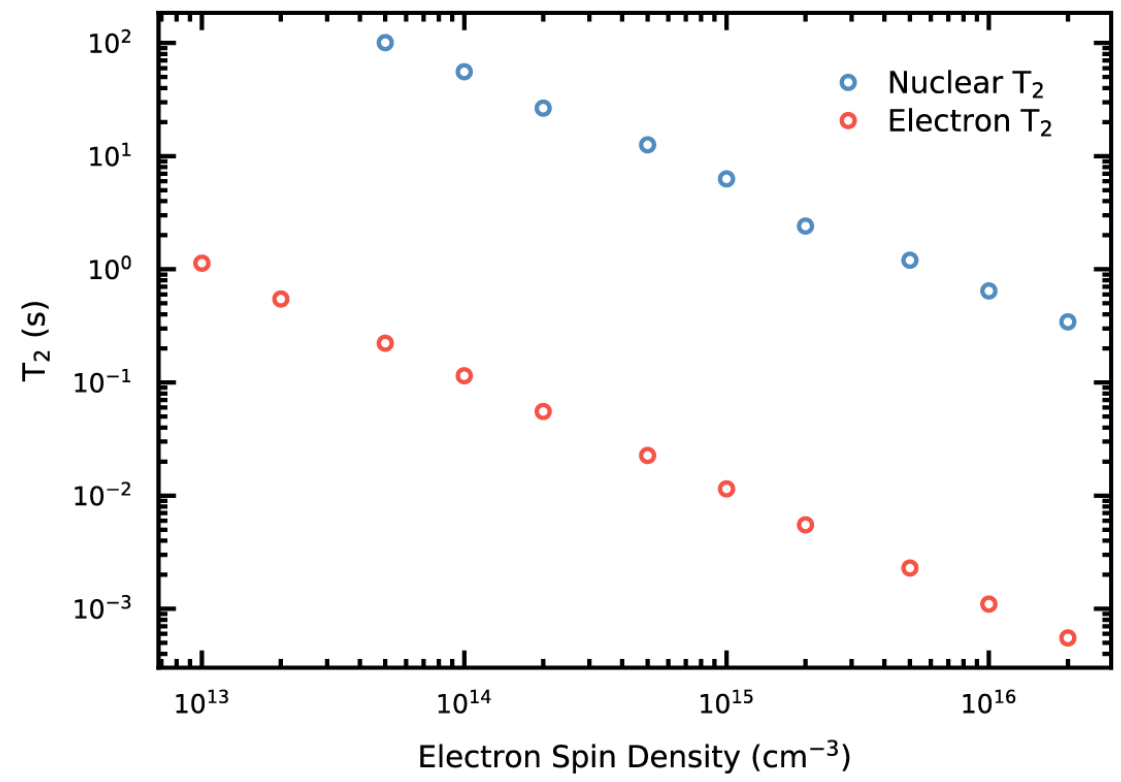

FIG 25. Electron and nuclear $T_2$ vs electron spin density. CCE-calculated Hahn echo $T_2$ of the $VV^0$ electron spin and $^{29}Si$ nuclear spin as a function of electron spin density.

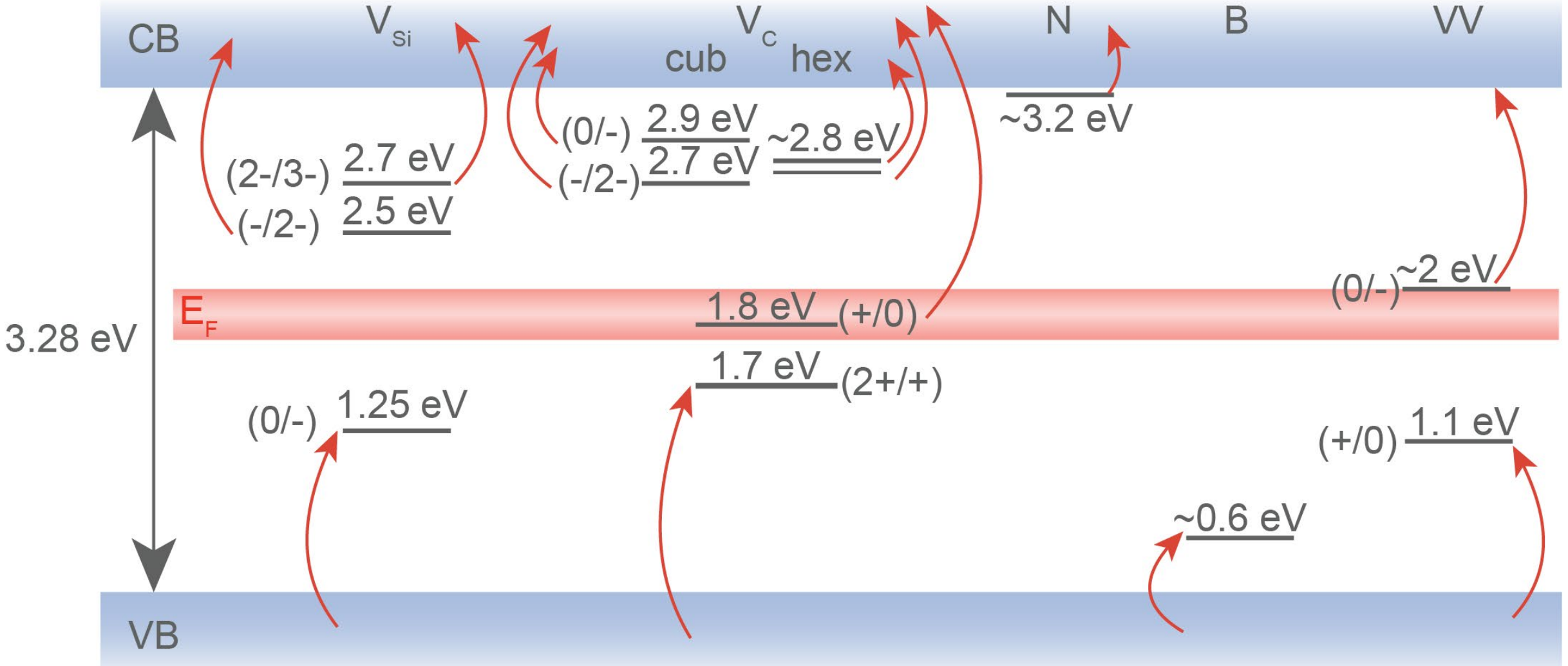

FIG 26. Band diagram of SiC and charge state of traps. The 705 nm illumination depopulates and stabilizes certain charge states of different deep and shallow traps within the band of SiC. The red band is the suggested Fermi level from the diode that stabilizes $VV^0$ and other traps to a certain charge state.

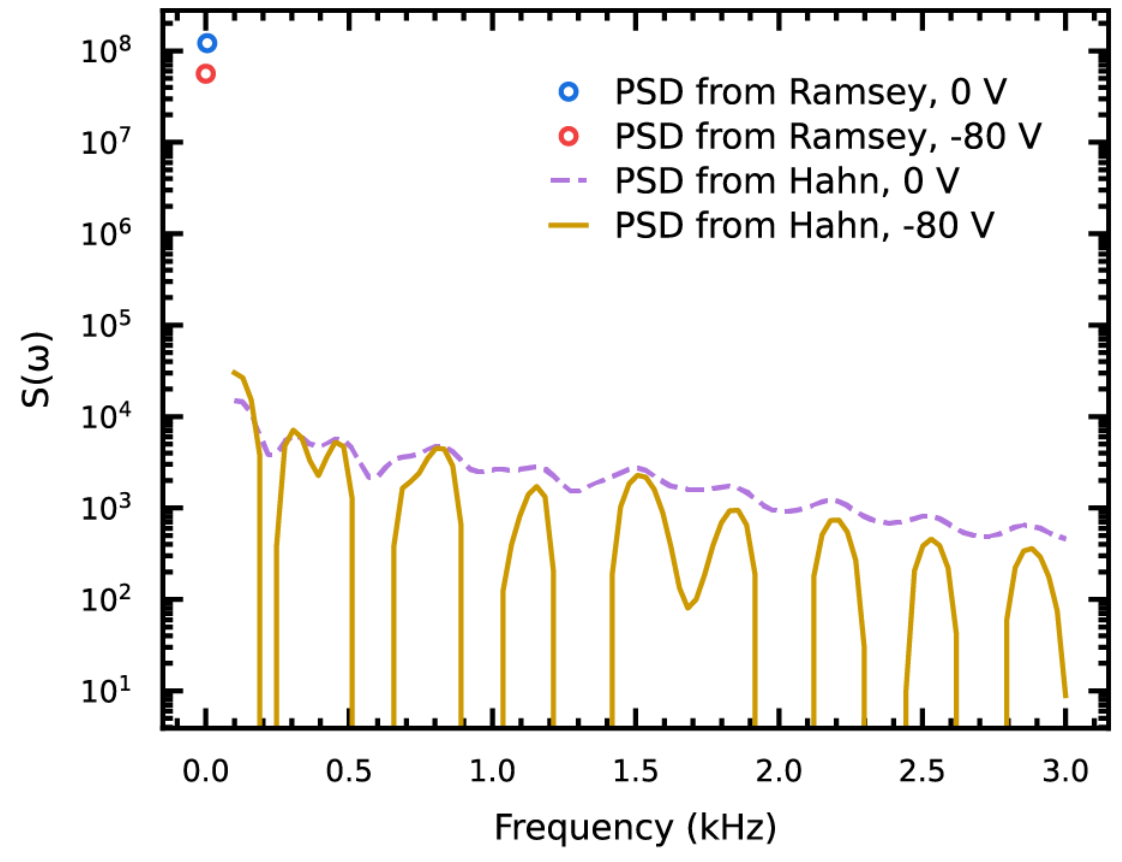

FIG 27. Reconstructed noise spectrum. Depleted and undepleted noise spectrum from Ramsey and Hahn echo measurements of the electron spin. The dips and non-smooth features from the -80 V Hahn echo PSD are artifacts from the numerical fitting algorithm.

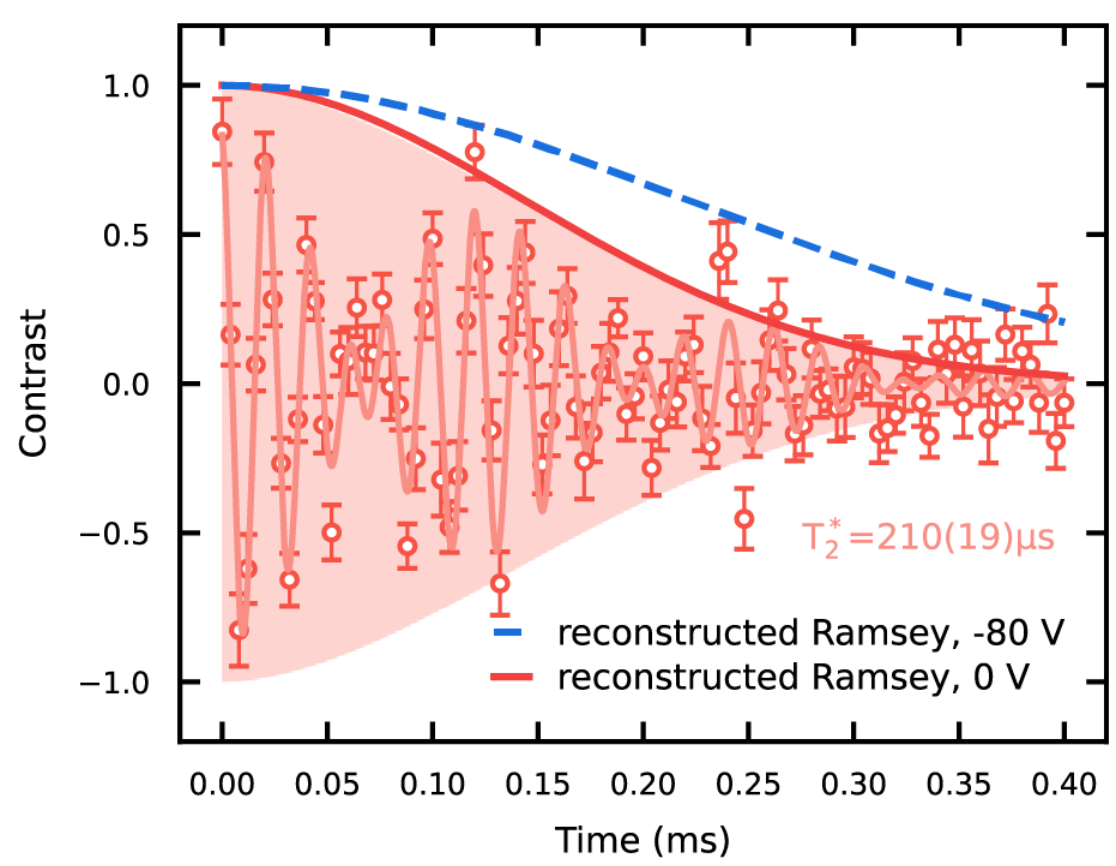

FIG 28. $VV^0$ Ramsey decay as predicted from the reconstructed noise spectrum. The $VV^0$ Ramsey decay predicted from the noise spectrum is plotted with the experimentally measured Ramsey decay in the undepleted regime.

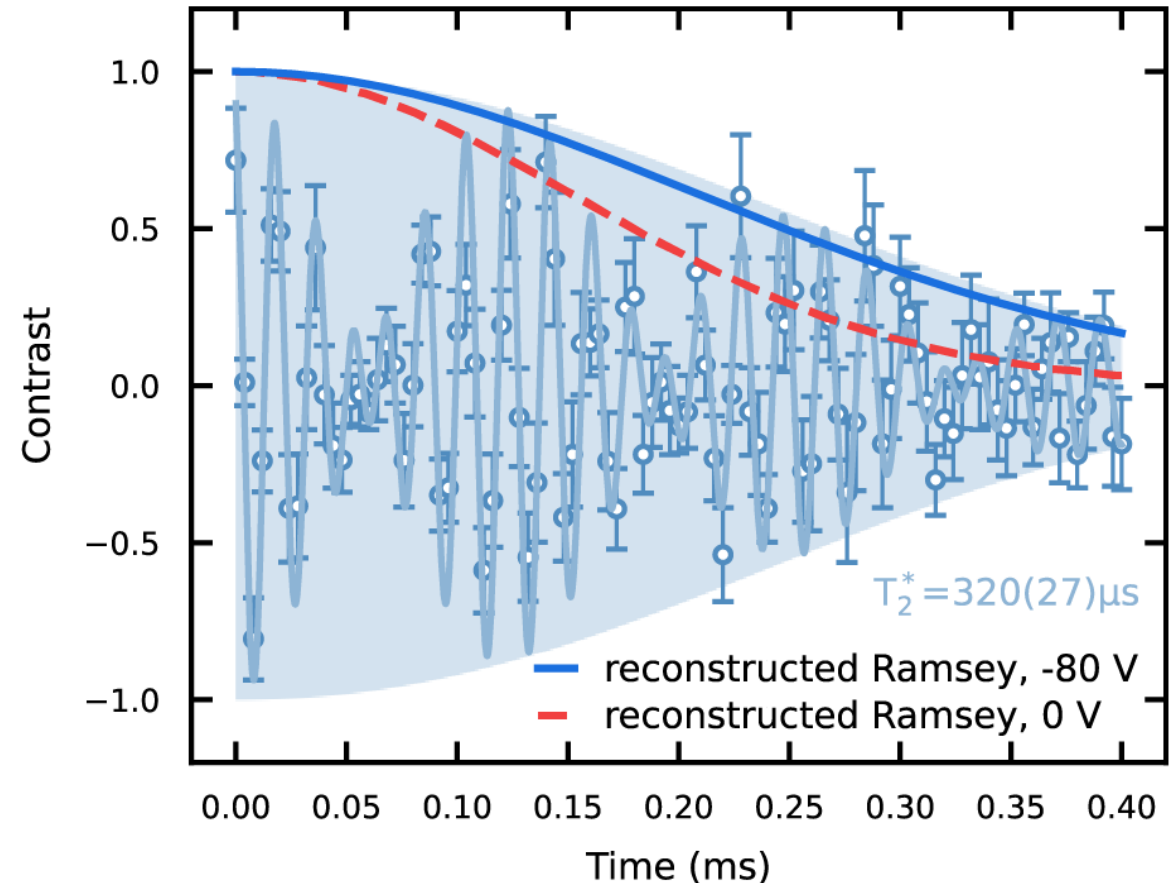

FIG 29. $VV^0$ Ramsey decay in the depletion regime as predicted from the reconstructed noise spectrum. The $VV^0$ Ramsey decay predicted from the noise spectrum is plotted with the experimentally measured Ramsey decay in the depleted regime.